\documentclass[11pt,a4paper]{revtex4}
\usepackage{amsmath}
\usepackage{graphicx}
\usepackage{fancyhdr}
\usepackage{calc}
\usepackage{amssymb}
\usepackage{setspace}
\usepackage{amsfonts}
\usepackage{commath}
\usepackage[innercaption]{sidecap}
\usepackage[framemethod=TikZ]{mdframed}
\usepackage{wrapfig}
\usepackage{hyperref}
\usepackage{parskip}
\usepackage{tabulary}

\usepackage{fullpage}
\usepackage{amsthm}
\usepackage{textcomp}
\usepackage{mathrsfs}
\usepackage{booktabs,hyperref}
\hypersetup{colorlinks=true}
\usepackage{verbatim}
\usepackage{tikz}
\usetikzlibrary{positioning}
\usepackage{ifthen}
\usepackage{booktabs}

\newcommand{\Rom}[1]{\expandafter\@slowromancap\romannumeral #1@}
\makeatother
{ \normalfont} 

\expandafter\def\expandafter\normalsize\expandafter{%
	\normalsize
	\setlength\abovedisplayskip{0pt}
	\setlength\belowdisplayskip{5pt}
	\setlength\abovedisplayshortskip{0pt}
	\setlength\belowdisplayshortskip{5pt}
}
\renewcommand{\baselinestretch}{1}
\definecolor{Gray}{gray}{0.75}

\newmdenv[backgroundcolor=Gray, leftmargin = 0pt, rightmargin = 0pt, linewidth = 0pt, roundcorner = 2 pt, innerleftmargin=5pt, innerrightmargin=5pt, innertopmargin=5pt, innerbottommargin=5pt]{Frame}

\begin{document}

\title{Alternating quarantine for sustainable epidemic mitigation}
\author{Dror Meidan$^{1}$, Nava Schulmann$^{2,3}$, Reuven Cohen$^{1}$, Simcha Haber$^{1}$, Eyal Yaniv$^{4}$, Ronit Sarid$^{5}$ \& Baruch Barzel$^{1,6,*}$}
\affiliation{
\begin{enumerate}
\item
Department of Mathematics, Bar-Ilan University, Ramat-Gan, Israel
\item
Department of Mechanical Engineering, Politecnico di Milano, Milan, Italy
\item
MIMESIS, Inria, Strasbourg, France
\item
Graduate School of Business Administration, Bar Ilan University, Ramat-Gan, Israel
\item
Faculty of Life Sciences \& Institute of Nanotechnology and Advanced Materials, Bar Ilan University, Ramat-Gan, Israel
\item
Gonda Multidisciplinary Brain Research Center, Bar-Ilan University, Ramat-Gan, Israel
\end{enumerate}
\begin{itemize}
\item[\textbf{*}]
\textbf{Correspondence}: baruchbarzel@gmail.com
\end{itemize}
}

\begin{abstract}
\textbf{Absent a drug or vaccine, containing epidemic outbreaks is achieved by means of social distancing, specifically mobility restrictions and lock-downs. Such measures impose a hurtful toll on the economy, and are difficult to sustain for extended periods. As an alternative, we propose here an alternating quarantine strategy, in which at every instance, half of the population remains under lock-down while the other half continues to be active, maintaining a routine of weekly succession between activity and quarantine. This regime affords a dual partition:\ half of the population interacts for only half of the time, resulting in a dramatic reduction in transmission, comparable to that achieved by a population-wide lock-down. All the while, it enables socioeconomic continuity at $50\%$ capacity. The proposed weekly alternations also address an additional challenge, with specific relevance to COVID-19. Indeed, SARS-CoV-2 exhibits a relatively long incubation period, in which individuals experience no symptoms, but may already contribute to the spread. Unable to selectively isolate these invisible spreaders, we resort to population-wide restrictions. However, under the alternating quarantine routine, if an individual was exposed during their active week, by the time they complete their quarantine they will, in most cases, begin to exhibit symptoms. Hence this strategy isolates the majority of pre-symptomatic individuals during their infectious phase, leading to a rapid decline in the viral spread, thus addressing one of the main challenges in COVID-19 mitigation.}
\end{abstract}

\maketitle

\vspace{-5mm}
Battling the spread of SARS-CoV-2, most countries have resorted to social distancing policies, imposing restrictions \cite{Anderson2020}, from complete lock-downs, to severe mobility constraints \cite{Kraemereabb4218,Arenas2020.04.06.20054320,article,Gross2020.03.23.20041517}, gravely impacting socioeconomic stability and growth. Current observations indicate that such policies must be put in place for extended periods (typically months) to avoid reemergence of the epidemic once lifted \cite{Hellewell2020,Zhigljavsky2020.04.09.20059451,YaneerBarYam2020}. This, however, may be unsustainable, as individual social and economic needs will, at some point surpass the perceived risk of the pandemic \cite{Epstein2009}.

More broadly, these events have exposed our vulnerability - socially and economically - to the emergence of novel infectious pathogens \cite{Hacohen2019}, calling on us to design socioeconomically sustainable response protocols, in the absence of therapeutic interventions. We, therefore, examine here an \textit{alternating quarantine} (AQ) strategy, tailored and tested for our immediate threat of COVID-19, but equally relevant to other pandemic spreading scenarios.

\begin{figure}[ht]
\centering
\includegraphics[width=16cm]{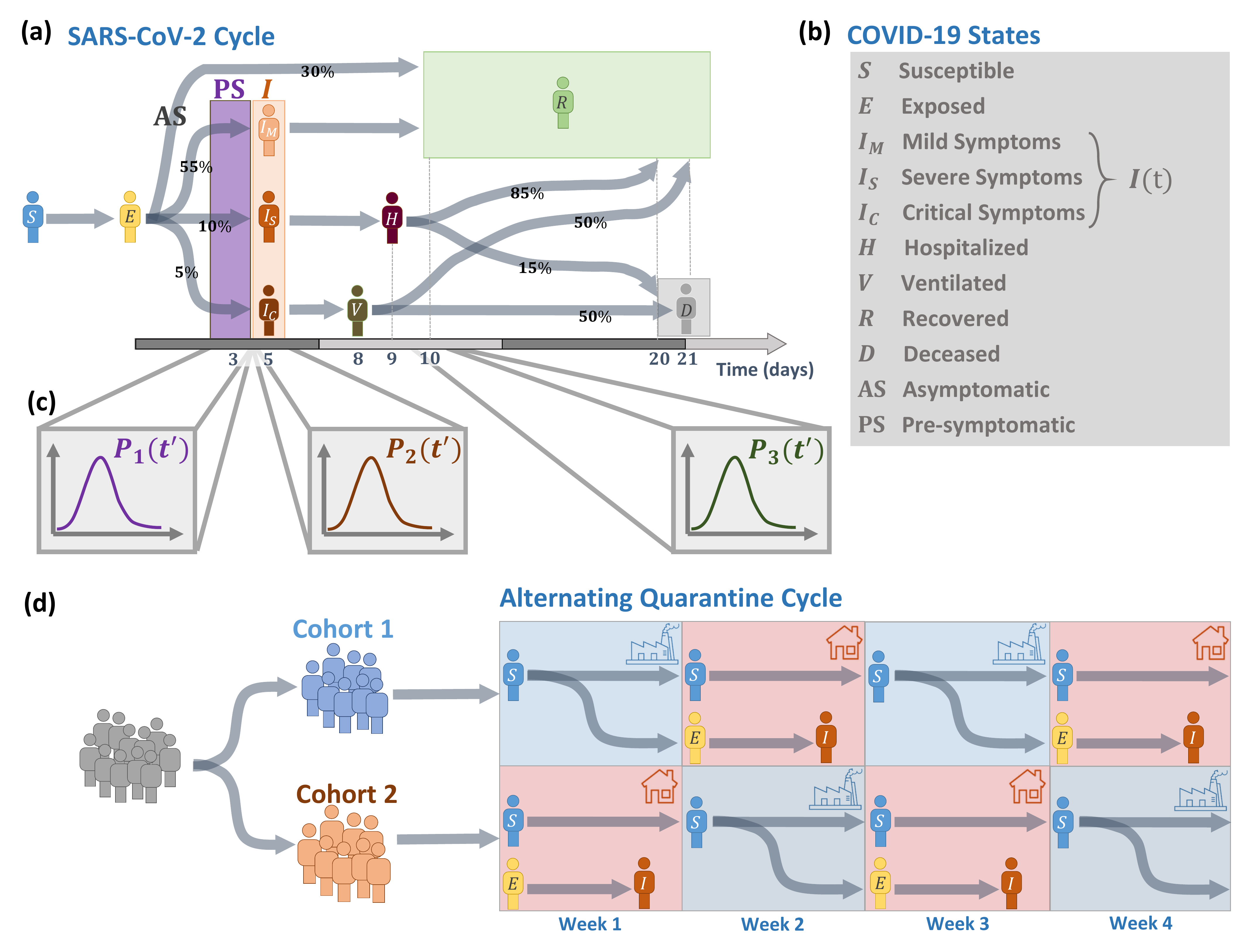}
\caption{\textbf{The cycles of SARS-CoV-2 and COVID-19 vs.\ those of the Alternating quarantine strategy}.\
(a)
We collected data on the transitions between the SARS-CoV-2 and COVID-19 states and constructed the characteristic disease cycle. Upon exposure ($E$) individuals enter an average $5$ day incubation period prior to developing symptoms - mild ($I_M$ at a rate of $55\%$), severe ($I_S$, $10\%$) or critical ($I_C$, $5\%$). The remaining $\sim 30\%$ are asymptomatic (AS). Infectiousness begins typically $3$ days after exposure for symptomatic carriers, and $4$ days for the asymptomatic (AS). The \textit{infection window} (violet) captures the invisible pre-symptomatic (PS) spreading phase, in which individuals are infectious, but lack symptoms. Upon the onset of symptoms, infected individuals are isolated and cease to infect others. Consequently, asymptomatic individuals have a longer infection window, which extends until their transition to $R$. As the disease progresses a fraction of the infected population may require hospitalization ($H$) or ventilation ($V$), leading, with some probability to mortality ($D$).
(b) The compartments of the COVID-19 cycle. We denote by $I(t)$ the unity of all \textit{symptomatic} individuals ($I = I_M + I_S + I_C$). This corresponds to the diagnosed case count in each country (Fig.\ \ref{Fitting}), which covers mainly the patients who exhibit symptoms.
(c)
While the illustrated cycle in (a) captures the average transition times between all states, in reality, some level of variability exists across the population. This is captured by the distribution $P_i(t^{\prime})$. For example the individual transition time from $E$ to PS, whose average is $3$ days, is extracted from $P_1(t^{\prime})$ (purple).
(d)
Alternating quarantine (AQ) splits the population into separate cohorts that alternate between periods of activity (going to work, blue) and inactivity (staying at home, red). Following their active week (week $1$) individuals in Cohort $1$ may become exposed (yellow), in which case they will sit out their suspected pre-symptomatic period at home (week $2$). By the end of their quarantine week they will likely develop symptoms (orange) and remain in isolation until their full recovery. Those who did not develop symptoms during their week of quarantine are most likely uninfected (blue) and can resume activity in their upcoming active shift (week $3$). Therefore the AQ cycle behaves as a \textit{ratchet}, consistently quarantining the invisible spreaders, and hence, removing, with each weekly succession, infectious individuals from the active population.}
\label{Illustration}
\end{figure}

The AQ strategy is based on two principles:\
(i) Complete isolation of symptomatic individuals and their household members \cite{Anderson2020};
(ii) Partitioning of the remaining households into two cohorts that undergo weekly successions of quarantine and routine activity. Other periodic cycles, \textit{e.g}., bi-weekly, or $5$ working days vs.\ $9$ quarantine days, may also be considered. The partition, we emphasize, must be at household level, guaranteeing all cohabitants are in the same cohort. Hence while Cohort $1$ remains active, Cohort $2$ stays at home and vice versa, ensuring little interaction between the cohorts (Fig.\ \ref{Illustration}d). This provides a highly efficient mitigation, alongside continuous socioeconomic productivity, in which half of the workforce remains active at each point in time.

The AQ strategy limits social mixing \cite{block2020social}, while providing an outlet for people to sustain their economic and social routines. Its efficiency is rooted in two independent mitigating effects:\

$\bullet$ \textit{Dual-partition of population and time} (Fig.\ \ref{Multiplier}).\ Splitting the population into two isolated cohorts reduces the number of infectious encounters. Indeed, classrooms, offices and public places operate at half their usual density, and hence, individuals interact with only half their usual contacts. On top of that, as each cohort is only active for half of the time, one week out of two, the infections within each cohort are further reduced, roughly by an additional factor of one half.

This dual-partition effect, relevant for any general pandemic, is reinforced by an additional factor, unique to COVID-19:\

$\bullet$ \textit{Synchronization with disease cycle} (Fig.\ \ref{Illustration}).\
AQ's weekly alternations treat one of the main obstacles for COVID-19 mitigation - the fact that while we isolate the symptomatic patients, exposed individuals become infectious a few days prior to the onset of symptoms \cite{Li2020,WHO2020,ferretti2020quantifying,ferguson2020impact,backer2020incubation,linton2020incubation} (Fig.\ \ref{Illustration}a). During this pre-symptomatic stage, they behave as \textit{invisible spreaders}, who continue to interact with their network, unaware of their potential infectiousness \cite{tao2020high,mizumoto2020estimating,pan2020asymptomatic,lu2020sars,al2020asymptomatic,colson2020children,song2020considerable,dong2020epidemiology}. To illustrate AQ's remedy, consider an individual in Cohort $1$ who was active during week $1$, and therefore might have been infected. This individual will soon enter their pre-symptomatic stage, precisely the stage in which they are \textit{invisible}, and hence contribute most to the spread. However, according to the AQ routine, they will be confined to their homes during week $2$, and consequently, they will be isolated precisely during their suspected pre-symptomatic period. If, by the end of week $2$ they continue to show no symptoms, most chances are that they are, in fact, healthy, and can, therefore, resume activity in week $3$ according to the planned routine. Conversely, if they do develop symptoms during their quarantine, they (and their cohabitants) must remain in isolation, similar to all symptomatic individuals. \textit{Hence, the weekly succession is in resonance with the natural} SARS-CoV-2 \textit{disease cycle} \cite{nason2020rapidly}, \textit{and in practice, leads to isolation of the majority of invisible spreaders. If implemented fully, it guarantees, in each bi-weekly cycle, to prune out the infectious individuals and sustain an active workforce comprising a predominantly uninfected population}.

Using COVID-19 as our test case, we examine below the performance of AQ, and discuss practical aspects pertaining to its implementation, from guidelines on how to partition the social network to the treatment of limited social compliance.

\vspace{3mm}

\newcommand{\m}[3]{#1_{#2#3}}
\newcommand{\av}[1]{\langle #1 \rangle}

\vspace{3mm}
{\color{blue} \Large \textbf{Analysis}}

{\color{blue} \textbf{Modeling the spread of SARS-CoV-2}}

\begin{figure}[ht]
\centering
\includegraphics[width=16cm]{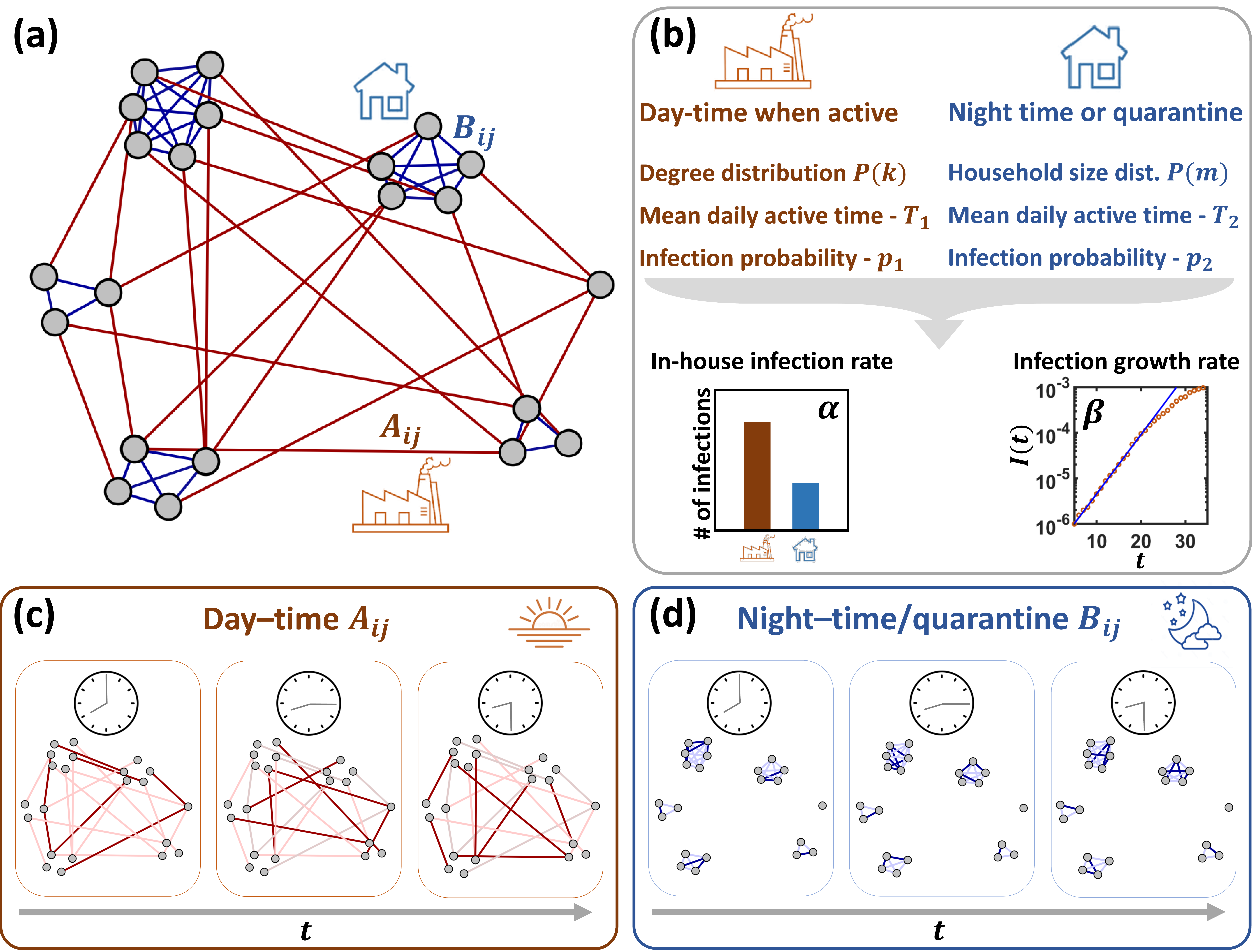}
\caption{\textbf{Modeling the spread of COVID-19}.\
(a) We constructed a \textit{social network} with two types of links:\ $\m Aij$ (red) are external links, representing out-of-home interactions at work, public places and social gatherings; $\m Bij$ (blue) capture cohabitant links, capturing separate households. In our setting the network includes $N = 10^4$ individuals split among $M = 4 \times 10^3$ households.
(b) The links in $\m Aij$ are \textit{active} primarily during the day-time, and only between individuals who are not under quarantine. The in-house connections $\m Bij$ are activated in the after-hours, or all day when under quarantine. Both networks are characterized by the degree/household size distributions $P(k)$ and $P(m)$. Their links exhibit sporadic instances of activity, capturing times when $i$ and $j$ are collocated, and therefore engage in potentially infectious interaction. Hence, they are characterized by two parameters:\ $T_1,T_2$ capture the typical time per day in which the links are active, and $p_1, p_2$ capture the probability of infection at each such instance of activity. This captures the fact that (i) cohabitant links are, typically, more active than social links ($T_2 > T_1$); (ii) when activated, cohabitant interactions are often more intimate and therefore potentially more infectious ($p_2 > p_1$). Together these six parameters, balancing the relative roles of $\m Aij$ and $\m Bij$ in the virus transmission, give rise to two observable parameters that help characterize the contagion:\ $\alpha$ in (\ref{Alpha}), quantifying the contribution of external (red) vs.\ in-house (blue) transmission to the spread; $\beta$ in (\ref{Beta}), describing the proliferation rate of the virus.
(c) We simulated social activity over a period of $T = 150$ days, at $15$ minute resolution. At each $15$ minute instance a fraction of the links are active (dark red), while the other remain idle. Transmission between $i$ and $j$ can only occur (with probability $p_1$) at times when $\m Aij$ is active (on average $T_1$ percent of the time). External links $\m Aij$ are active primarily during the day, excluding periods of quarantine.
(d) $\m Bij$ undergo a similar pattern of activity/idling, with parameters $T_2,p_2$, primarily in the evening/night-time, or all day during quarantine.
}
\label{Modeling}
\end{figure}

\textbf{Social network}.\ We constructed a population of $N \sim 10^4$ individuals, comprising $M = 4 \times 10^3$ separate households, and tracked their sequence of social interactions at $15$ minute resolution over the course of $T = 150$ days. These interactions are governed by two separate networks:\ day-time interactions at work, school or other public places are driven by the external network $\m Aij$. This represents an $N \times N$ network with degree distribution $P(k)$, designed to capture out-of-home social activity (Fig.\ \ref{Modeling}a, orange). In-house interactions, taking place predominantly during the after-hours, are governed by $\m Bij$, a network of $M$ isolated cliques, capturing households (Fig.\ \ref{Modeling}a, blue). The size of these cliques $m$ is extracted from the empirically obtained \cite{UNHouseholds} household size distribution $P(m)$.

To capture the temporal nature of the interactions, each link in $\m Aij$ and $\m Bij$ switches between periods of activity, \textit{i.e}.\ collocation of $i$ and $j$, vs.\ intermittent periods, in which the link remains idle. Infection between $i$ and $j$ may occur during the instances in which the $i,j$ link is active. These instances of activity/inactivity, extracted at random, represent the sporadic nature of human interaction, and allow us to track the viral spread under highly realistic conditions. To design a typical daily cycle, the external links $\m Aij$ are predominantly active during the day (Fig.\ \ref{Modeling}c), while in-house interactions occur primarily at the evening/night-time hours (Fig.\ \ref{Modeling}d). During quarantine, such as under the AQ routine, or if a household member $i$ exhibits symptoms, the relevant $\m Aij$ links remain idle, while $\m Bij$ becomes activated also during the day.

Each of the networks $\m Aij$ and $\m Bij$ is characterized by two independent parameters (Fig.\ \ref{Modeling}b). The first captures the probability of link-activation at each $15$ minute interval, determining the links' mean daily duration of activity. We denote this duration by $T_1$ for $\m Aij$ and $T_2$ for $\m Bij$. Realistically we expect $T_1 < T_2$, capturing the fact that cohabitants, when at home, spend more time in potentially infectious interactions than, \textit{e.g}., office-mates during work hours. The second parameter is the probability of transmission per interaction, set to $p_1$ and $p_2$ for $\m Aij$ and $\m Bij$ respectively. Also here we assume that typically $p_1 < p_2$, as in-house interactions, often between family members, are potentially more infectious than the social interactions of $\m Aij$. For example, parents tending to children or siblings interacting physically, are more likely to lead to infection, than co-workers sharing a meeting room.

\textbf{Disease cycle}.\ In Fig.\ \ref{Illustration}a we present the SARS-CoV-2 characteristic infection cycle. Upon exposure ($E$) individuals enter a pre-symptomatic period, which lasts, on average $\sim 5$ days, after which they begin to exhibit mild ($I_M$), severe ($I_S$) or critical ($I_C$) symptoms, leading to hospitalization ($H$), and in certain cases also to ventilation ($V$). Approximately $2$ days prior to the onset of symptoms the exposed individuals become infectious, hence, on average, the infectious phase begins $3$ days after initial exposure \cite{Li2020}. Spreading the virus continues until the onset of symptoms, at which point the infected individuals, together with their cohabitants, enter isolation and cease to contribute to the external spread ($\m Aij$). Of course, in-house transmission ($\m Bij$) continues also during home-isolation. A fraction of the exposed individuals remain asymptomatic (AS), and hence do not isolate, throughout their entire infectious period, beginning on average $4$ days posterior to exposure \cite{bar2020sars}. Hence, the symptomatic carriers spread the disease within an average window of $\sim 2$ days (purple), while the asymptomatic carriers continue to infect others until their immune response clears the virus.

These time-scales represent the \textit{average} infection cycle, which, in reality, may exhibit variability across the population \cite{ferretti2020quantifying,ferguson2020impact,backer2020incubation,linton2020incubation,li2020early,jiang2020does,bar2020sars,lauer2020incubation}. This is especially relevant regarding the time for the appearance of symptoms, which, if delayed beyond $1 - 2$ weeks, may lead to an infectious crossover between successive terms of activity, \textit{e.g.}, if a person is infected in week $1$, and then, lacking symptoms, resumes activity in week $3$ (see Fig.\ \ref{Illustration}d). Therefore, for each of the relevant time-scales, \textit{e.g.}, the time from exposure to infectiousness, or the time to develop symptoms, we consider not just the \textit{average}, but the \textit{complete distribution} across the population (Fig.\ \ref{Illustration}c). For example, the probability density function $P_1(t^{\prime})$ captures the fraction of exposed individual who exhibit symptoms within $t \in (t^{\prime}, t^{\prime} + \dif t^{\prime})$ days from exposure. Similarly, $P_2(t^{\prime})$ characterizes the transition times between exposure and asymptomatic infectiousness. The broader are $P_i(t^{\prime})$, the greater is the individual variability in transition times between the different disease states. Here we extract $P_i(t^{\prime})$ from a Weibull distribution, as indicated for SARS-CoV-2 \cite{linton2020incubation,Lauer2020,backer2020incubation}, and observed also for other infections of Corona type viruses \cite{bar2020sars}; see Supplementary Sections 1.3 and 4.1.

\textbf{Characterizing the spread} (Fig.\ \ref{Modeling}b).\ Taken together, our modeling framework is designed to capture the spread of SARS-CoV-2 in a detailed fashion, therefore, characterized by an array of relevant parameters. The majority of these parameters can be extracted from empirical data. For example, the transition rates of Fig.\ \ref{Illustration}a's disease cycle, or the household size distribution $P(m)$, helping us structure $\m Bij$ - are all empirically accessible. The remaining parameters $P(k), T_1, T_2, p_1, p_2$, however, are unknown. Hence, we examine different spreading scenarios to examine our strategy's sensitivity to discrepancies is these parameters. For instance, we consider both a random $\m Aij$, for which $P(k)$ is bounded, and a scale-free $\m Aij$, where $P(k)$ is fat-tailed (Supplementary section 2). Similarly, we assign different values to $T_1$ and $T_2$ to examine variable balance between in-house and external transmission.

Once these unknown parameters are set, they help characterize the spread along two dimensions (Fig.\ \ref{Modeling}b):\

$\bullet$ \textit{The growth rate} $\beta$, quantifies the level of spread by tracking the initial exponential proliferation of infections

\begin{equation}
I(t) \sim e^{\beta t},
\label{Beta}
\end{equation}

observed at the early stages of the outbreak. Here $I(t) = I_M(t) + I_S(t) + I_C(t)$ is the number of symptomatic infected individuals. The greater is $\beta$ the more rapid is the spread, hence $\beta$ increases with $T_1,T_2,p_1$ and $p_2$, all of which contribute to the infectiousness of the disease. The density of the network, \textit{i.e}.\ its average external degree $\av k$ and household size $\av m$ also both positively contribute to $\beta$, as they allow more potential infectious interactions. This parameter is directly related to the disease's doubling time \cite{bar2020sars,li2020early} via $T_{\rm Double} \sim \beta^{-1} \ln 2$.

$\bullet$ \textit{The in-house infection rate} $\alpha$, captures the balance between the contribution of $\m Bij$ vs.\ $\m Aij$ to the spread. To quantify this balance we track all instances of transmission $\theta_{\rm Tot}$, and extract $\theta_{\rm In}$, which counts only the cases transmitted via $\m Bij$ links, \textit{i.e}.\ in-house. We then measure the in-house infection rate as

\begin{equation}
\alpha = \dfrac{\theta_{\rm In}}{\theta_{\rm Tot}}
\label{Alpha}
\end{equation}

namely the fraction of transmissions that occurred at home.

Similarly to $\beta$, the parameter $\alpha$ is also dictated by the network parameters. A large $T_2$ and $p_2$ will favor in-house transmissions, contributing to $\alpha$, whereas increasing $T_1, p_1$ will strengthen the role of external transmissions. The network structure also factors in through the average number of external links $\av k$, decreasing $\alpha$, vs.\ the typical household size $\av m$, which increases it. This parameter is especially meaningful in the context of quarantine-based strategies, which, by design, suppress only $\theta_{\rm Tot}$, and therefore become less effective when $\theta_{\rm In}$ is large. Indeed, no matter how strict the quarantine is, it cannot prevent the secondary transmission between household members encapsulated within $\theta_{\rm In}$. In fact, it often increases these in-house infections, as it forces cohabitants to remain close for extended periods. Consequently, a large $\alpha$ can potentially hinder the effectiveness of quarantine in general, and of AQ in particular.

To summarize, in our modeling we vary the \textit{implicit} model parameters $T_1,p_1,T_2,p_2,P(k)$, to scan an array of potentially relevant scenarios. Once setting these parameters, we use them to extract two \textit{observable} parameters $\alpha$ and $\beta$, that directly characterize the nature of the contagion. The severity of the spread is quantified by $\beta$, and the role of in-house transmission is captured by $\alpha$. The mapping between the model parameters and the observable $\alpha,\beta$ is explained in Supplementary Sections 1.4.

{\color{blue} \textbf{Projecting the spread of SARS-CoV-2}}

\begin{figure}[ht]
\centering
\includegraphics[width=16cm]{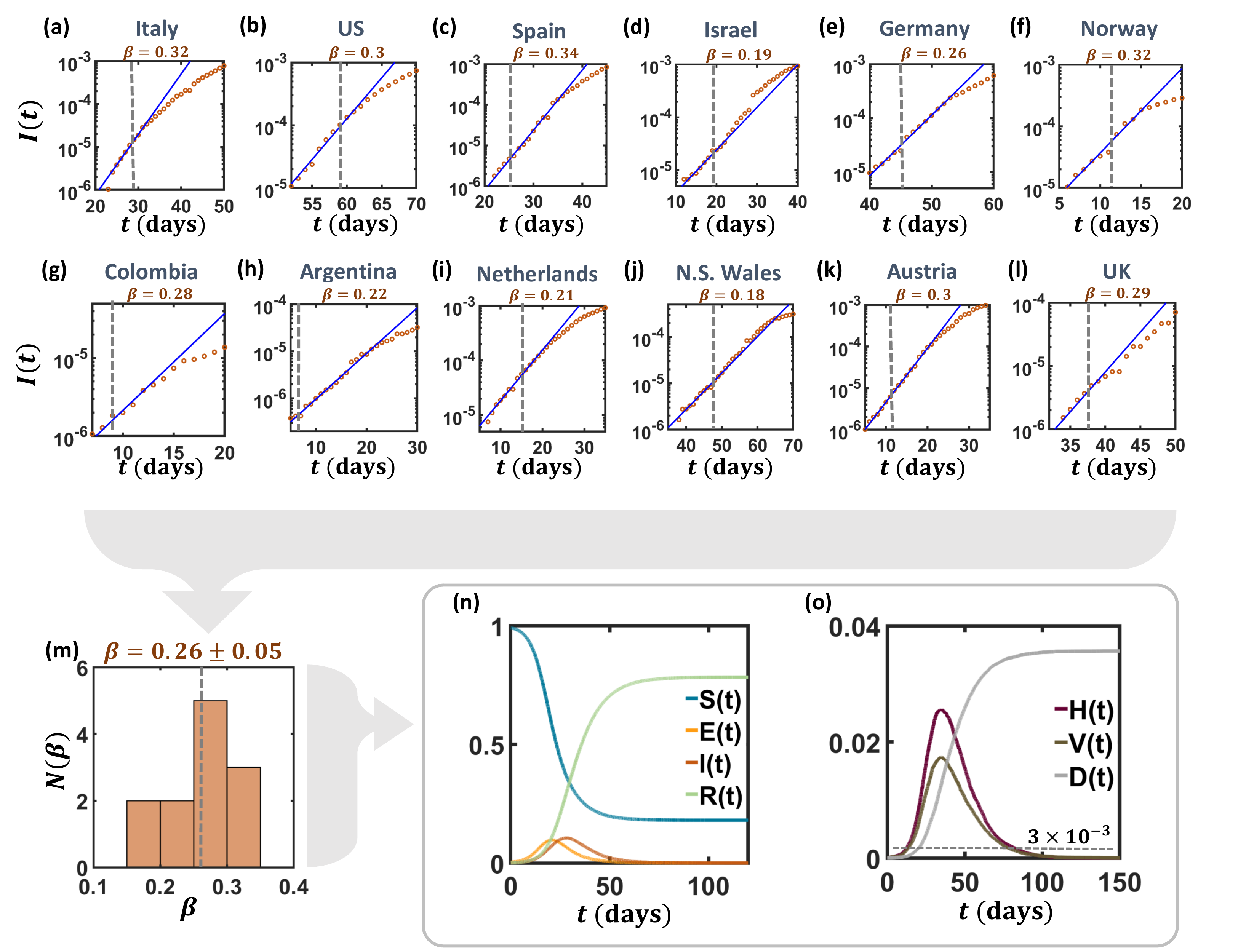}
\caption{\textbf{Extracting SARS-CoV-2 infection rate}.\
(a) - (l)
We collected data \cite{JohnHopkins2020} on the infection levels $I(t)$ vs.\ $t$ (orange circles) in $12$ different countries. Fitting $I(t)$ to an exponential of the form (\ref{Beta}) we evaluate $\beta$ in each of these countries (blue solid lines). Such exponential growth typically continues for a period of several days posterior to the instigation of a mitigation policy (grey dashed-lines). We, therefore, used only the data up to three days after the implementation of social distancing to evaluate $\beta$ (Supplementary Section 4.2).
(m)
Histogram of inferred $\beta$ values across the $12$ countries. Infection rates are distributed around an average of $\beta = 0.26$. Hence, in our simulations we tune the parameters to obtain this growth rate under the absence of all prophylactic measures. In reality, standard behavioral practices, such as personal hygiene or avoidance of physical contact, may push $\beta$ to lower values. To capture this, in our simulations we incorporate three scenarios: worst case - $\beta \approx 0.26$, intermediate case - $\beta \approx 0.2$ and best case - $\beta  \approx 0.15$.
(n)
Taking $\beta \approx 0.26$ we simulated the projected evolution of the COVID-19 pandemic \textit{a l\`{a}} Figs.\ \ref{Illustration} and \ref{Modeling}, without any preventive measures.
(o)
We focus on three crucial parameters that characterize the severity of the projected spread:\ mortality $D(t)$ (grey), hospitalization level $H(t)$ (purple) and the number of individuals requiring ventilation $V(t)$ (brown). Absent any intervention, $H(t)$ exceeds, by a large margin, the average national hospitalization capacity (dashed grey line), estimated at $3 \times 10^{-3}$ \cite{Worldbank}.
Results represent an average over $20$ stochastic realizations.
}
\label{Fitting}
\end{figure}

To evaluate $\beta$ for the unmitigated COVID-19 spread we collected data on the evolution of the epidemic in $12$ different countries \cite{JohnHopkins2020}, and examined $I(t)$ at the early stages of the spread, prior to the implementation of social distancing policies (Fig.\ \ref{Fitting}a-l). Fitting to an exponential growth of the form (\ref{Beta}) we evaluate $\beta$ in each country, obtaining a mean growth rate of $\beta = 0.26$ days$^{-1}$, an estimate congruent with other independent evaluations \cite{bar2020sars,li2020early}. Setting, as a baseline $\alpha = 0$ (to be changed below) we can now obtain a projection of the expected evolution of the epidemic (Fig.\ \ref{Fitting}n). We also track the expected fraction of hospitalized ($H(t)$) and ventilated ($V(t)$) individuals (Fig.\ \ref{Fitting}o), which, we find, exceed, by a significant margin, the average national hospitalization capabilities \cite{Worldbank}. The expected mortality is captured by $D(t \rightarrow \infty)$ reaching, absent any mitigation efforts, a level of $\sim 3\%$ (grey).

Next, we examine the behavior of COVID-19 under AQ, together with other relevant strategies.

\vspace{3mm}
{\color{blue} \Large \textbf{Mitigation}}

\begin{figure}[ht]
\centering
\includegraphics[width=16cm]{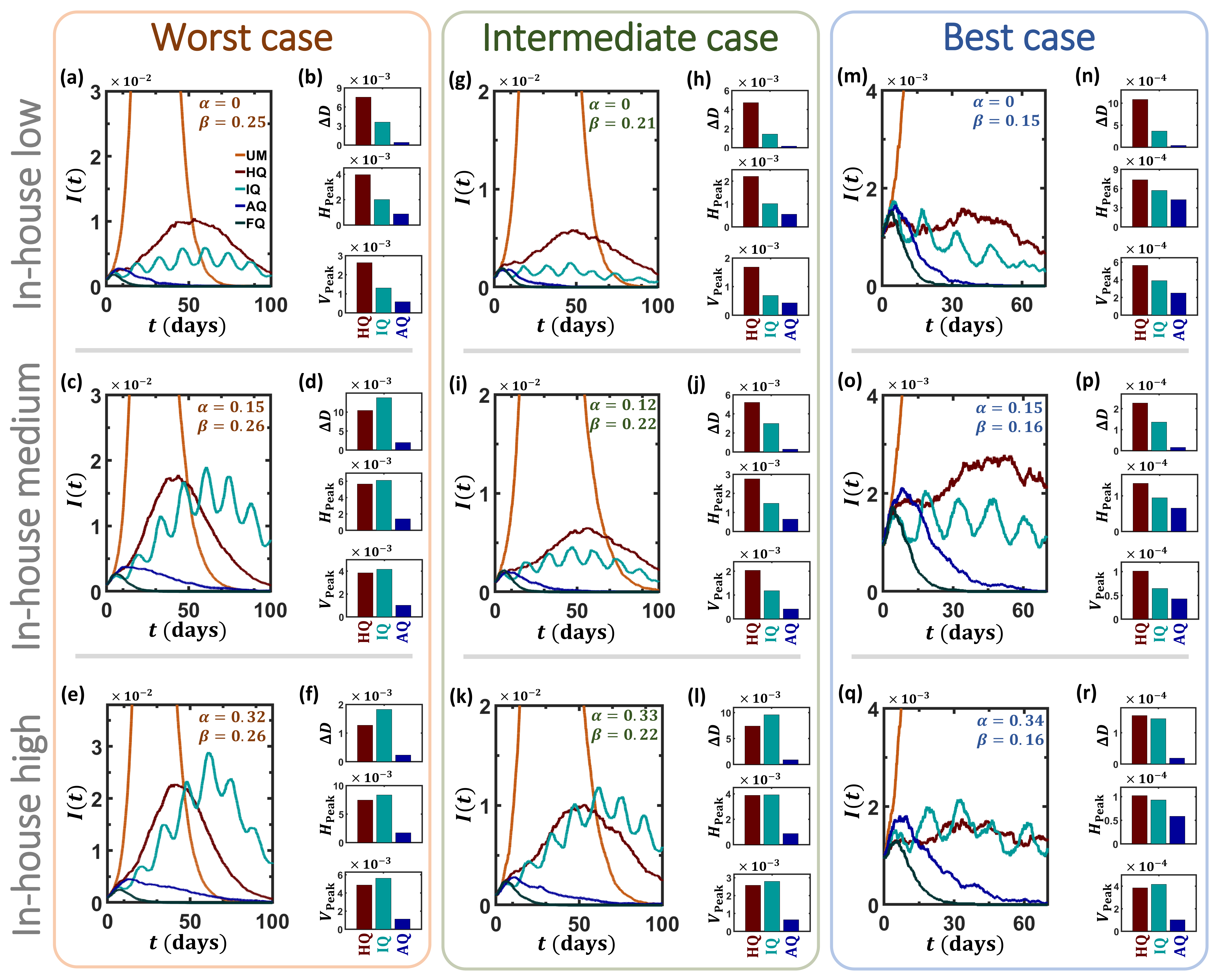}
\caption{\textbf{The impact of Alternating quarantine}.\
(a)
The infection $I(t)$ vs.\ $t$ of the unmitigated epidemic (UM, orange), as obtained under $\beta = 0.25$ and $\alpha = 0$. This represents a worst-case scenario, where $
beta$ is taken from the unmitigated spreading data, \textit{i.e}.\ lacking prophylactic practices. At $t = t_0$ we instigate four competing mitigation strategies:\ Full quarantine (FQ, grey), Alternating quarantine (AQ, blue), Intermittent quarantines (IQ, turquoise) and Half quarantine (HQ, red). We find that, barring the \textit{idealized} FQ, AQ provides the most efficient mitigation, outperforming IQ and HQ.
(b)
We used three performance measures to rate mitigation efficiency:\ Residual mortality $\Delta D$ (\ref{DeltaD}), peak hospitalization $H_{\rm Peak}$ (\ref{HPeak}) and peak ventilation $V_{\rm Peak}$. In all three indicators, AQ provides the best outcomes by a significant margin, as compared to IQ and HQ.
(c) - (f)
To examine the role of in-house transmission, we further tested all strategies under medium ($\alpha = 0.15$) and high ($\alpha = 0.32$) levels of household infections.
(g) - (l)
We repeated the same experiment, this time under a $\sim 20\%$ lower $\beta$, capturing the potential effect of complementary prophylactic measures, such as mask wearing or hygienic behavior.
(m) - (r)
In our best case scenario $\beta$ is further reduced by an additional $\sim 20\%$ factor, this time to
$\beta \approx 0.15$. Together, our analysis scans the space of infection ($\beta$) and in-house transmission ($\alpha$) rates, covering a range of potentially relevant conditions for COVID-19 mitigation. We find that, under \textit{all} conditions, AQ consistently outperforms all contending strategies, providing mitigation that is closest to FQ.
Results represent an average over $20$ stochastic realizations. Mitigation is initiated at time $t_0$, set to be the time when $I(t)$ exceeds $10^{-3}$ (Supplementary Section 1.2). In our simulations, the external network $\m Aij$ was taken to be an Erd\H{o}s-R\'{e}nyi random graph with average degree $\av k = 15$; in Supplementary Section 2 we also examine the case of a scale-free $\m Aij$, obtaining similar results. Note that $\alpha$ and $\beta$ are controlled indirectly, first through the model parameters ($P(k),P(m),T_1,T_2,p_1,p_2$) as illustrated in Fig.\ \ref{Modeling}b, then extracted from the observed stochastic simulation results via Eqs.\ (\ref{Beta}) and (\ref{Alpha}). Consequently, we cannot control with accuracy the precise values of these parameters. We, therefore, observe slight discrepancies between the different panels, \textit{e.g}., $\beta = 0.25$ in panel (a) vs.\ $0.26$ in panels (c) and (e). For full size image see Page 21.
}
\label{Results}
\end{figure}

\begin{figure}[ht]
\centering
\includegraphics[width=12cm]{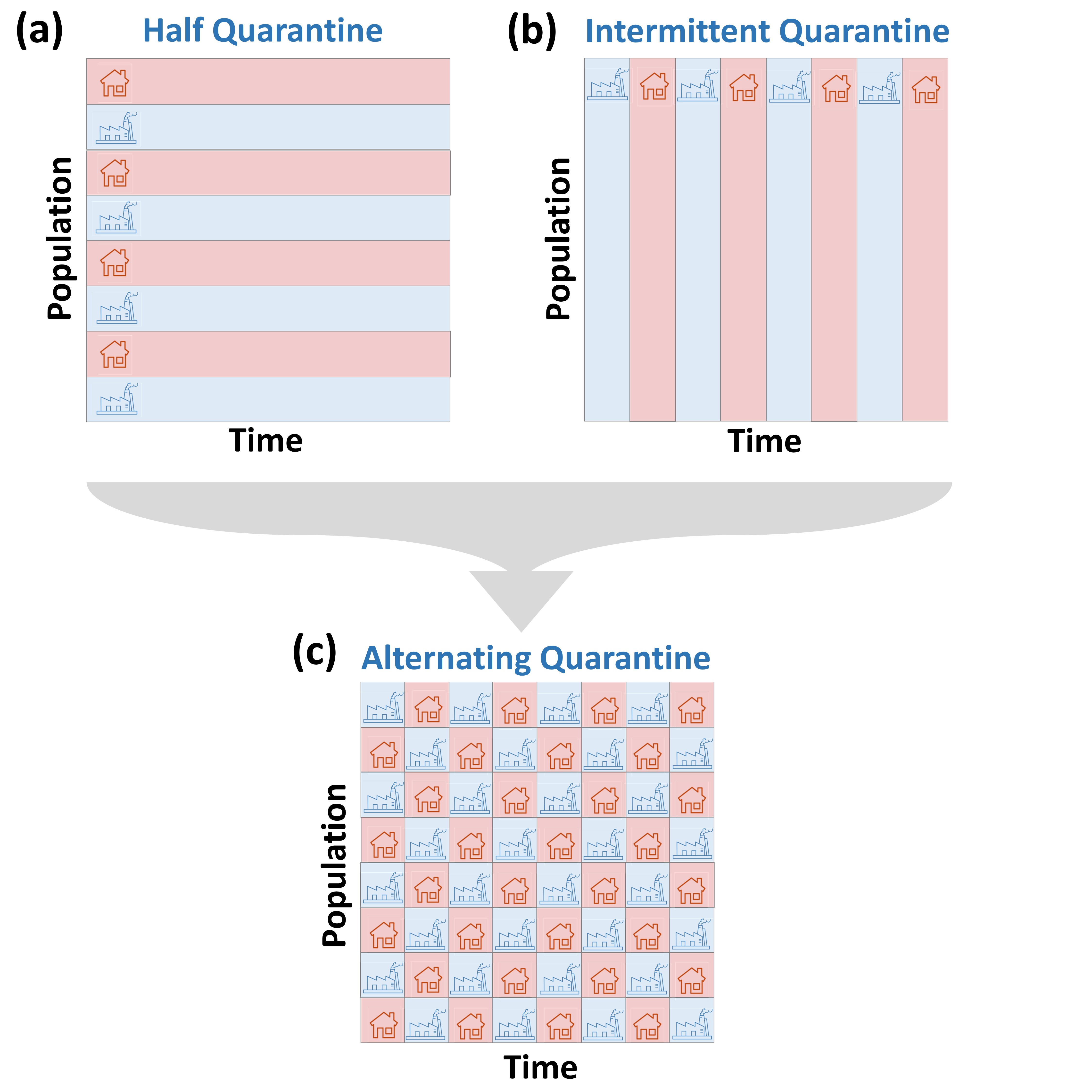}
\caption{\textbf{The multiplicative effect of Alternating quarantine}.\
We consider three strategies - all allowing socioeconomic activity (blue) vs.\ quarantine (red) at half capacity.
(a)
The Half quarantine strategy reduces infection by diluting the active \textit{population}, hence decreasing the rate of infectious interactions.
(b)
Intermittent quarantines achieve a similar outcome by diminishing the duration of activity, hence reducing the \textit{time} of infectious interactions.
(c)
Our alternating quarantine combines both effects:\ on the one hand interactions are limited to individuals within each cohort - diluting the population. On the other hand these cohorts experience intermittent cycles of work/home - diminishing interaction duration. The result is a $\sim 4$-fold reduction in transmission, alongside a mere $50\%$ reduction in socioeconomic activity.
}
\label{Multiplier}
\end{figure}

To examine the impact of our proposed strategy we track the evolution of $I(t) = I_M(t) + I_S(t) + I_C(t)$. First we allow the disease to spread unmitigated (Fig.\ \ref{Results}a, orange, UM), then at time $t = t_0$ (Supplementary Section 1.2) we instigate our response. Examining four relevant mitigation strategies, we establish a basis upon which to evaluate AQ's performance.

{\color{blue} \textbf{Full quarantine - FQ}} (Fig.\ \ref{Results}a-f, grey).\
This represents the \textit{theoretically ideal} response, in which all out of home activity is ceased. The external links $\m Aij$ become inactive and only in-house transmission ($\m Bij$) remains, until these secondary infections are also exhausted and the spread reaches a halt. To capture the effect of this in-house perpetuation of the disease we consider several scenarios, from vanishing in-house transmission (Fig.\ \ref{Results}a, $\alpha = 0$) to extreme levels of household infections (\ref{Results}e, $\alpha = 0.32$). As expected, absent in-house transmission, FQ eradicates the disease extremely efficiently, within $\sim 3$ weeks. As $\alpha$ is increased, FQ, expectantly, shows a slight decline in efficiency \cite{YaneerBarYam2020}. Of course, such perfect \textit{air-tight} quarantine is unrealistic, however, it is useful in the present context, as it provides a baseline for comparison, indeed, setting the bounds for a \textit{perfect} mitigation.

{\color{blue} \textbf{Alternating quarantine - AQ}} (Fig.\ \ref{Results}a-f, blue).\
We now examine the AQ strategy. At $t = t_0$ we partition the households into two equal groups, cohorts $1$ and $2$, and have them alternating successively in a bi-weekly cycle of quarantine, \textit{i.e}.\ only $\m Bij$ is active, and regular socioeconomic activity, namely $\m Aij$ \textit{and} $\m Bij$ are active. We find, again, that $I(t)$ decays exponentially, albeit at a slower rate, as compared to the \textit{prefect} FQ. \textit{The crucial point, however, is that this decay is now observed, despite the fact that $50\%$ of the population remains continuously active}.

For comparison, we consider two natural alternatives to AQ, both designed to sustain socioeconomic activity at a $50\%$ rate:

{\color{blue} \textbf{Intermittent quarantines - IQ}} (Fig.\ \ref{Results}a-f, turquoise).\
In this strategy \cite{Karin2020.04.04.20053579} society as a whole enters a periodic cycle of active vs.\ quarantined phases, namely the entire population transitions in unison between staying at home and going to work. Originally proposed in the format of a $4:10$ periodicity, \textit{i.e}.\ $4$ days of activity separated by $10$ days of quarantine, here we examine its performance under a $7:7$ cycle, to be congruent with our implementation of AQ. We find that IQ is significantly less effective than AQ, leading not only to higher peak infection, but also to a substantially longer time to return to normalcy.

{\color{blue} \textbf{Half quarantine - HQ}} (Fig.\ \ref{Results}a-f, red).\
Another mitigation alternative that allows a $50\%$ active workforce is based on a selective quarantine, in which only $50\%$ of the population partakes in socioeconomic activities, while the remaining half is instructed to stay at home. HQ suppresses the rate of infection by reducing social interactions, \textit{i.e}.\ $\m Aij$, by a factor of roughly one half. Our simulation results indicate, however, that, similarly to IQ, this reduction is insufficient. Indeed, $I(t)$ continues to proliferate significantly beyond manageable levels, once again, failing to mitigate the disease.

While the majority of infected individuals exhibit mild or no symptoms, a certain percentage may experience severe complications, leading to hospitalization or ventilation, and in some cases to mortality (Fig.\ \ref{Illustration}a). Our mitigation strategy focuses on these undesired paths within the infection track - namely, we aim to decrease mortality $D(t)$, and ensure that at their peak, $H(t)$ and $V(t)$ do not exceed the national hospitalization and ventilation capabilities. To test this we measured the residual mortality

\begin{equation}
\Delta D = D_{\rm S}(t \rightarrow \infty) - D_{\rm FQ}(t \rightarrow \infty)
\label{DeltaD}
\end{equation}

where $D_{\rm S}(t \rightarrow \infty)$ is the long term mortality under strategy $\rm S$, \textit{e.g.}, IQ or AQ, and $D_{\rm FQ}(t \rightarrow \infty)$ is the expected mortality under FQ. Indeed, within the framework of quarantine-based strategies, $D_{\rm FQ}(t \rightarrow \infty)$ represents inevitable deaths, rooted in infections that occurred prior to our response, and hence $\Delta D$ captures the \textit{additional} mortality, that our mitigation failed to prevent. In Fig.\ \ref{Results}b,d,f we measure $\Delta D$ under IQ (turquoise), HQ (red) and AQ (blue). The AQ advantage is clearly visible, saving significantly more lives than the competing strategies.

To examine the impact of each strategy on the severe and critical patients, we measure

\begin{equation}
H_{\rm Peak} = \max_{t = t_0}^\infty H(t),
\label{HPeak}
\end{equation}

capturing the peak hospital occupancy \textit{after} instigating our response. While IQ and HQ fail to bring $H_{\rm Peak}$ within capacity ($\sim 3 \times 10^{-3}$), AQ succeeds in sustaining a leveled occupancy. Similar results are also obtained for $V_{\rm Peak} = \max_{t = t_0}^\infty V(t)$.

\vspace{2mm}
\begin{Frame}
Taken together, we find that AQ provides the most efficient mitigation, bringing us closest to the ideal performance of FQ, without fully shutting down the economy. To understand the origins of the observed AQ advantage, we first consider its alternatives, IQ and HQ. The common root of both strategies is that they reduce the level of interaction by a factor of one half. IQ achieves this by decreasing the interaction \textit{duration}; HQ accomplishes this by diluting the interacting \textit{population}. In this sense, the strength of AQ is that it benefits from both factors (Fig.\ \ref{Multiplier}):\ partitioning the population into cohorts ensures that only half are active at all times - similar to HQ. Yet, the weekly alternations ensure that each cohort remains active only half the time - similar to IQ. This dual partition further reduces infectious interactions without increasing the socioeconomic toll. On top of that, AQ is also tailored specifically to the COVID-19 time-scales, with its weekly periodicity, roughly in-phase with the natural $\sim 5$ day cycle of incubation and pre-symptomatic infection (Fig.\ \ref{Illustration}d). \textit{The result is an effective force multiplier, allowing the same amount of net activity - $50\%$ - but with a dramatically enforced mitigation effect}.
\end{Frame}

\vspace{3mm}
{\color{blue} \Large \textbf{Synergistic measures}}

Our analysis, up to this point, assumed a \textit{worst case scenario}, in which, aside from our mitigation strategy (AQ, IQ or HQ), all other disease parameters remain unchanged. In reality, however, in addition to AQ, or any other strategy for that matter, we can expect, at the least, that standard prophylactic behaviors will continue to be practiced. Indeed, personal hygiene, face-masks and contact avoidance can reduce infections significantly, without taking any toll on the economy. Therefore, in practice, the infection rate $\beta = 0.26$, inferred in Fig.\ \ref{Fitting} from the early, pre-mitigation stages of the epidemic, will likely be reduced as we gradually adapt to a prophylactic routine. We, therefore, examine the performance of the different mitigation strategies also under a reduced $\beta$, capturing the synergistic effect offered by such practices. In the \textit{intermediate case} we set $\beta \approx 0.21$, a $\sim 20\%$ reduction in the rate of infection (Fig.\ \ref{Results}g - l), and as our \textit{best case} scenario, we examined $\beta \approx 0.15$, capturing a $\sim 40\%$ drop in infectiousness (Fig.\ \ref{Results}m - r). Under these more favorable conditions, AQ's performance approaches even closer to the ideal FQ (\textit{e.g.}, Fig.\ \ref{Results}g or m), providing a dramatic reduction in mortality and hospitalization.

More generally, our AQ strategy can, and should, be reinforced by other complementary policies, to ensure mitigation success, from avoiding social gatherings to establishing isolation facilities, with the purpose of reducing in-house transmission. As a specifically relevant example, we consider, in Supplementary Section 3, the selective protection of vulnerable populations, addressing a crucial aspect of COVID-19, whose impact on the elderly or on individuals with co-morbidities, is disproportionately more severe \cite{Mallapaty2020,Bonafe2020,Xie2020}.

All of these policies can be instigated alongside, rather than instead, of AQ. One may also consider alternative periodic cycles \cite{Karin2020.04.04.20053579}. For instance, a $5:9$ cycle, in which the active shifts last only $5$ days. In this version of AQ, society enters a routine in which each cohort is allowed a $5$ day work-week, then observes population-wide quarantine over the weekend. Such adaptations will further improve the performance of AQ beyond its already established effectiveness.

\begin{figure}[ht]
\centering
\includegraphics[width=15cm]{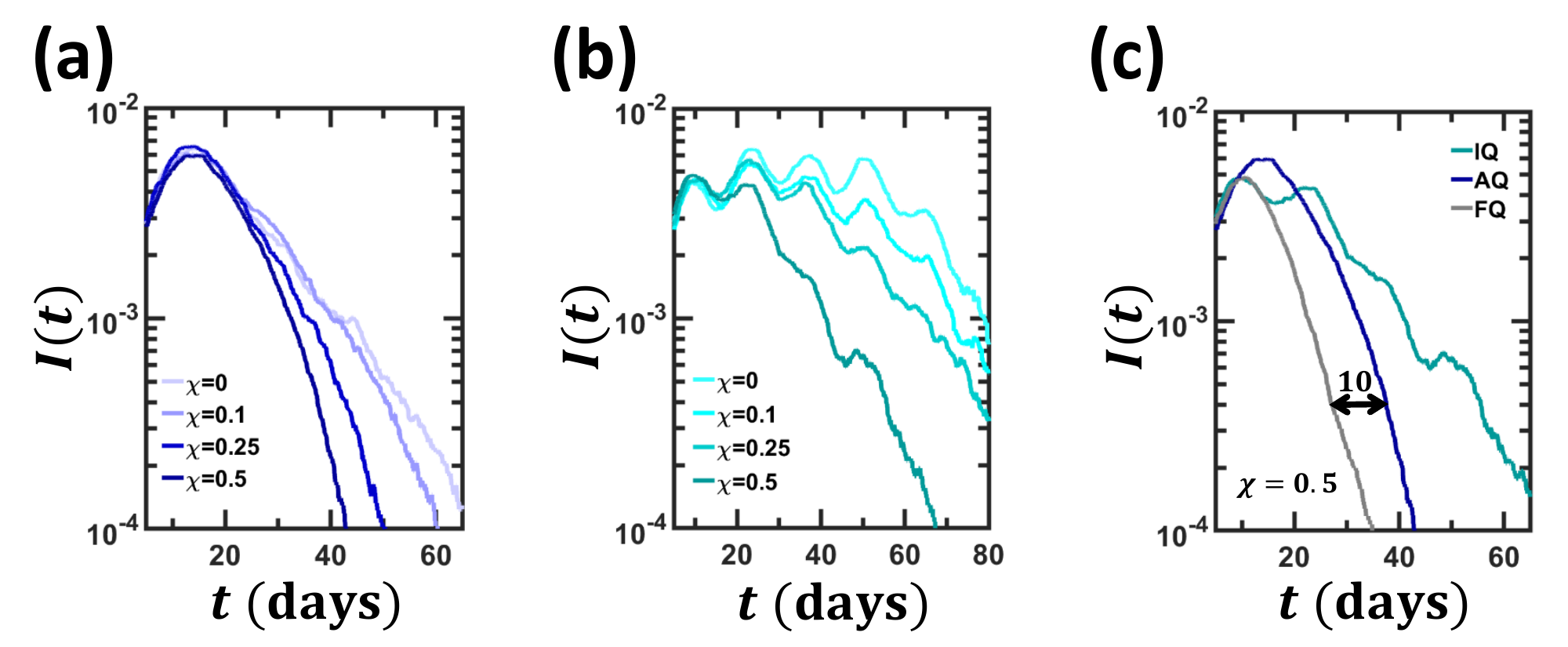}
\caption{\textbf{Testing the population before resuming activity}.\
(a)
We combine AQ with testing at a weekly capacity of $\chi$. Directing all tests towards the quarantined cohort at each week, we prohibit individuals who tested positive (and their households) from resuming activity. Unsurprisingly, as $\chi$ increases (darker) we observe an enhanced mitigation, thanks to the systematic pruning of infected individuals from the active cohort.
(b)
Testing also improved the performance of Intermittent quarantines (IQ). As opposed to AQ, in IQ, as the entire population transitions from quarantine to activity in unison, the testing cannot be selectively directed to the quarantined cohort, but rather spread evenly across the entire population.
(c)
In the limit where $\chi \rightarrow 0.5$, a capacity to screen $50\%$ of the population within one week, AQ becomes extremely efficient, thanks to its natural partitioning of the population. The entire quarantined cohort can be tested, and within $1 - 2$ weeks AQ has almost no out of home infections. Indeed, we observe that AQ follows a similar decay as the Full quarantine (FQ, grey), albeit with a $10$ day delay, capturing roughly $1-2$ testing cycles. In contrast, IQ, lacking such partitioning of the population, exhibits a more minor benefit under the same testing capacity.
Simulations represent an average over $20$ stochastic realizations. The in-house infection rate was set to the intermediate level $\alpha \approx 0.15$.
}
\label{Testing}
\end{figure}

{\color{blue} \textbf{Population-wide testing}} (Fig.\ \ref{Testing}).\
Thanks to the synchronization with the disease cycle, each weekly quarantine filters out a fraction of the infected individuals. It is therefore natural to reinforce this filtering with systematic testing of the quarantined cohort before they resume activity. If an individual is detected positive, their entire household must remain in isolation until all members are cleared. To examine this effect we added a component of random testing to both AQ and IQ. Given limited resources, we assume a testing capacity of a $\chi$-fraction of the population per week. As expected, the greater is $\chi$ the more effective is our mitigation (Fig.\ \ref{Testing}a,b).

The crucial point, however, is that AQ's breakdown of the population into separated cohorts provides an intrinsic advantage. Indeed, testing is most effective when conducted on the quarantined cohort, whose state is \textit{frozen} during the week. One can then spread the testing across the entire week, and detect infected individuals before they return to activity. Hence, the fact that one only needs to focus on half of the population at a time, enhances the effectiveness of such a testing policy. To understand this, consider the case where $\chi \approx 0.5$, namely we have the capacity to test $50\%$ of the population within a single week. Under these conditions, thanks to AQ's partitioning, one can simply invest \textit{all} tests in the inactive cohort, then resume activity in week $2$ with a guaranteed clean workforce.

To examine this advantage we focus specifically on the case where $\chi = 0.5$. We apply the tests selectively to the quarantined cohort in each shift, which, indeed, constitutes roughly half of the population (minor discrepancies arise due to statistical variations, and uneven household sizes). Within a one week cycle we arrive at an almost $100\%$ uninfected active workforce, after which the only bottleneck for the decay of $I(t)$ is the residual in-house infections. In that sense, after approximately $1 - 2$ weeks, AQ becomes as effective as FQ. Indeed, Fig.\ \ref{Testing}c shows that AQ (blue) exhibits the same rate of $I(t)$ decay as FQ (grey), albeit at a $10$ day delay, precisely the predicted $1 - 2$ weeks. Hence, extensive testing provides a crucial complement to AQ, potentially achieving FQ mitigation efficiency, without crippling socioeconomic activity.

\vspace{3mm}
{\color{blue} \Large \textbf{Alternating vs.\ population-wide quarantine}}

The proven advantages of AQ indicate that it is not merely an alternative to IQ or HQ, both partial quarantine strategies, but may actually be confronted against a population-wide quarantine (PWQ). For example, a PWQ at rate $\eta$ requires an $\eta$ fraction of the population to continuously practice quarantine, hence in HQ we have $\eta = 50\%$ and in FQ we have $\eta = 100\%$.

Intuitively, one would expect a PWQ with $\eta > 50\%$ to be more effective than AQ, both in terms of mitigation - isolating larger parts of the population, as well as in terms of implementation - not having to resolve between the two cohorts. Our analysis, however, indicates that AQ has crucial advantages on both fronts. The implementation challenge of PWQ is that it requires people to stay at home for a period of several weeks, in order for the mitigation to take effect. For example, in Fig.\ \ref{Results}a we found that an $\eta = 100\%$ perfectly implemented quarantine (FQ), which is, indeed, a theoretical construction only, still required several weeks to achieve a significant gain over the disease \cite{YaneerBarYam2020}. Under these conditions, one cannot implement a truly complete lock-down. Essential services, supply chains and some parts of the market must remain active, since households cannot retain supplies and remain self-sufficient for such extended periods. Therefore, a practical PWQ can at most be implemented at a level of $\eta = 70 - 75\%$ \cite{khadilkar2020optimising}.

\begin{figure}[ht]
\centering
\includegraphics[width=15cm]{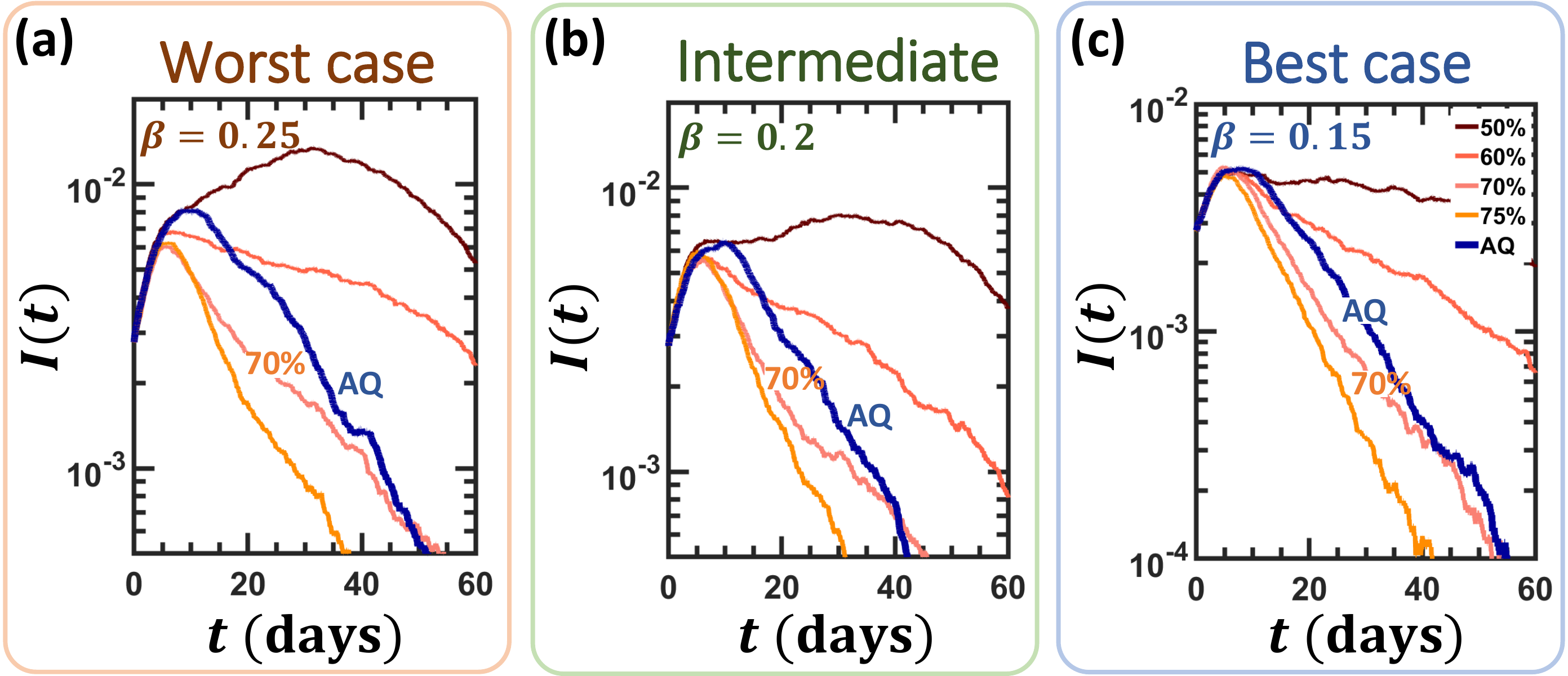}
\caption{\textbf{Alternating quarantine vs.\ population-wide quarantine}.\
(a)
Infection level $I(t)$ vs.\ $t$ as obtained for $\beta = 0.25$ under Alternating quarantine (AQ, blue). We also examined population-wide quarantines at $\eta = 50, 60, 70$ and $75\%$ levels (red to yellow). Despite having half of the population active at all times, AQ's mitigation is comparable to that of a $70\%$ population-wide quarantine. Hence, instead of an extremely hurtful socioeconomic shutdown of $70\%$, nearing the practical upper bound of social distancing policies, AQ offers a similar outcome under a significantly reduced socioeconomic price-tag.
(b) - (c)
Similar results are observed also under our intermediate ($\beta = 0.2$) and best case ($\beta = 0.15$) scenarios.
Simulations represent an average over $20$ stochastic realizations. The in-house infection rate was set to the medium level $\alpha \approx 0.15$. Here $\m Aij$ is an Erd\H{o}s-R\'{e}nyi random graph with $\av k = 15$, similar results under for a scale-free $\m Aij$ appear in Supplementary Section 2.
}
\label{PWQ}
\end{figure}

In contrast, the AQ scheme requires individuals to isolate only for a single week at a time. Hence, the quarantined cohort can truly enter, for just one week, a complete lock-down regime, in which they avoid purchasing supplies or any other services. Consequently, under AQ, while a larger part of the population is active at all times, the quarantined cohort, can sustain a much stricter lock-down routine. As a result not only is the economy more productive, with $50\%$ of the population continuously active, but the mitigation outcome is also comparable, and under some conditions even superior. To demonstrate this, in Fig.\ \ref{PWQ} we examine the impact of PWQ, imposed at a level of $\eta = 50, 60, 70$ and $75\%$ (red to yellow). AQ, we find, is roughly the equivalent of a $70\%$ lock-down (blue). Note that $\eta = 70 - 75\%$ represent the practical upper bound for any realistic PWQ. Yet whereas PWQ at such levels severely compromises the economy and imposes significant social and psychological stress, AQ accomplishes a similar effect, while sustaining a productive economy, and allowing a manageable routine for the individual.

\vspace{3mm}
{\color{blue} \Large \textbf{Implementation}}

\textbf{Partitioning}.\ The AQ strategy works best when the two cohorts are fully separated, lacking all forms of cross-group infection. The partition should, therefore, be implemented at a household level, ensuring all cohabitants are in the same activity/quarantine cycle. A simple way to achieve this is to base the partition on a person's living address. This provides an additional benefit, in the case of apartment buildings, as neighbors, who risk cross-infection through shared building facilities, are included in the same cohort. Each individual/household will be informed by their local authority of their quarantine schedule, and in parallel, employers will be instructed to resume their activity in shifts, with only half the workforce at a time. Businesses will be held legally liable and incur fines in case of violation.

Instances of conflict between a person's assigned shift and their personal/employer's specific requirements will be resolved on a \textit{case by case} basis - all while strictly adhering to the household-based partition. To encourage cooperation, and to ensure AQ's smooth implementation, it is best to be as flexible as possible in responding to all individual requests. The resulting cohorts, after accommodating such requests, will likely deviate from an exact balanced cut, however, the crucial point is, that the partition need not be perfect, as, indeed, the cohorts must be decoupled, but not necessarily equal in size. Therefore, there is much room to address specific constraints or special needs, thus easing the psychological and socioeconomic stress as much as possible. See \textbf{Box I} for a smooth partitioning scheme.

\begin{figure}[ht]
\centering
\includegraphics[width=15cm]{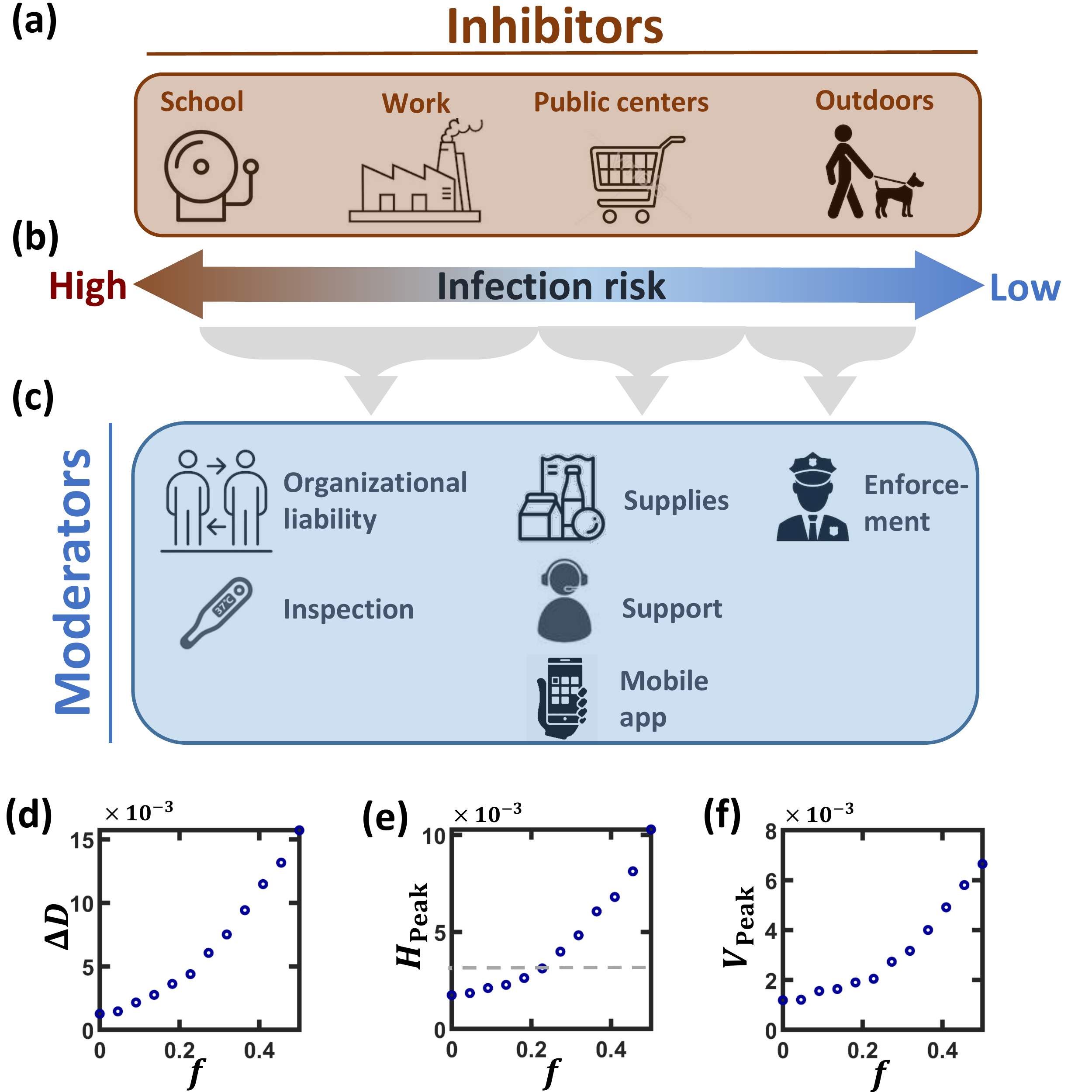}
\caption{\textbf{Driving social conformity for Alternating quarantine}.\
(a)
We identify four needs that inhibit potential cooperation:\ child care arrangements, work, purchasing supplies or services and outdoor activities.
(b)
Infection risk is highest under extensive and continuous interactions, such as in school or at work, and least significant during open-air activities, such as strolling or exercising. We therefore focus on moderators mainly for the first three inhibitors.
(c)
To enhance social compliance we seek moderators that encourage conformity in lieu of coercive enforcement:\
\textit{School and work}.\ Due to their liability, schools and workplaces will be naturally prohibited for the quarantined cohort, as both will be required to abide by the AQ routine, and therefore will not admit workers or students of the inactive cohort. In addition, routine inspections for symptoms will expose potential defectors who wish to conceal their infection. \textit{Public centers}.\ We consider three moderators to deter individuals from seeking services or supplies:\ (i) instruct the population to obtain sufficient supplies in advance for a single week; (ii) establish a support network in case of unexpected needs; (iii) create a mobile app confirming an individual's cohort ($1$ or $2$), that must be displayed upon entry to public centers. Outdoor activity could be moderated by enforcement, however, since it poses little infection risk, we believe such activity should, in practice, be ignored.
(d)
The residual mortality $\Delta D$ vs.\ the fraction of defectors/essential workers $f$, as obtained for the intermediate scenario $\beta = 0.20, \alpha = 0.15$. We find that AQ is robust under imperfect implementation, allowing to sustain a $\sim 20\%$ violation, either via formal exemption or by defection.
(e) - (f)
Similar results are obtained for $H_{\rm Peak}$ and $V_{\rm Peak}$; the average estimated national hospitalization capacity is indicated by the grey dashed-line.
}
\label{Implementation}
\end{figure}

\textbf{Social compliance}.\
To engage the population towards cooperation \cite{TAWSE2019}, the first step is to communicate the rationale behind AQ, its potential effectiveness, and the individual compliance required for its rapid success. This appeals to people's intrinsic motivation \cite{BAUMANN2017}, a crucial component of conformity, but often also insufficient due to the \textit{tragedy of the commons}. We therefore list the \textit{drivers}, that enhance people's desire to cooperate, vs.\ the \textit{inhibitors}, that stand in their way \cite{VROOM1964,MUSAWIR2020}, and set appropriate \textit{moderators} to enforce the drivers and suppress the inhibitors (Fig.\ \ref{Implementation}).

\textbf{Inhibitors} (Fig.\ \ref{Implementation}a).\
During its lock-down cycle, the quarantined cohort is required to stay at home for one week, indeed, a challenge, however, being limited in time, it is significantly less stressful than an extended several week quarantine. We identify four motivators to violate the quarantine:\ \textit{Business} - going to work, \textit{Schooling} - arrangements for child care, \textit{Services and supplies} - visiting public market places or service centers, and \textit{Outdoors} - exercise or strolling with children or pets. Of these, the latter, being in the open, is least risky, and also practically unavoidable, as young children and pets require routine outdoor activity. We, therefore, focus on moderators especially for the first three inhibitors.

\textbf{Moderators} (Fig.\ \ref{Implementation}c).\
While cooperation can be achieved via coercion, \textit{e.g.}, law enforcement, it is most effectively garnered by creating supporting frameworks. For example, in the AQ framework, defection for business and schools is simply not possible. Indeed, since businesses are legally required to divide their workforce into shifts, one cannot go to work out of cycle. Similarly, schools will not admit children who are not in the presently active cohort. Therefore, the main challenge is to deter violators from visiting public places for supplies or services. This can be achieved by (i) instructing the population to prepare in advance for a full week of isolation; (ii) establishing a logistic and psychological support network to aid citizens who encounter unexpected needs; (iii) creating a dedicated app to issue exit permits only to members of the active cohort. The app in (iii) will not violate citizen privacy in any way, but only indicate if the device holder is in Cohort $1$ or $2$. Residents will be asked to present their app to enter shopping centers or public institutions. This can be done in addition to testing for symptoms, as already practiced in many countries.

Together, the proposed moderators create a framework that not only diminishes incentives for defection, \textit{e.g}., by logistically supporting the isolated cohort, but also eliminates the means, as, indeed, aside from daily outdoor strolling, practically all other out of home activities are automatically barred by the AQ framework itself. The strength of this implementation plan is that it achieves this without coercion, namely that almost no enforcement via authorized forces against individuals is required, maintaining a level of trust between citizen and government and securing personal freedoms. To complete the plan, at the end of the isolation week, all isolated residents will be required to report their health status via the app. Those who report symptoms, as well as their cohabitants, will remain at their \textit{stay-home} status, going into isolation until their verified recovery.

{\textbf{Defectors and essential workers}} (Fig.\ \ref{Implementation}d-f).\
Despite this detailed implementation plan, some level of violation of the AQ regime is unavoidable. This is either due to partial compliance, \textit{i.e}.\ defectors, or because certain individuals hold essential positions and cannot leave their post for an entire week. Therefore, we now introduce a fraction $f$ of continuously active individuals, defectors or essential workers, who remain active at all times, both during their open shift as well as when their cohort is under quarantine. This $f$-fraction is extracted from the non-symptomatic ($S,E,I_{AS}$) or mild symptomatic ($I_M$) population, who may conceal their state. Excluded, however, are individuals experiencing severe symptoms ($I_S,I_C,H,V$) who, of course, remain isolated. Measuring our performance indicators, $\Delta D$ (\ref{DeltaD}), $H_{\rm Peak}$ (\ref{HPeak}) and $V_{\rm Peak}$, we find that AQ can sustain defection/exemption up to $f \sim 0.2$, a $20\%$ non-conformity level. Beyond that, we observe a significant decline in the strategy's performance.

\vspace{3mm}
{\color{blue} \Large \textbf{Discussion}}

The efficiency of the AQ strategy is rooted in three principles:\
(i) Partitioning the population into two cohorts reduces the volume of infectious interactions, comparable to a $50\%$ quarantine (HQ).
(ii) Working in weekly succession reduces the total duration of interaction within each cohort, similar to intermittent quarantines (IQ).
Combining these two factors together, allows a similar net volume of socioeconomic activity as in any of these strategies, HQ or IQ, \textit{but with a multiplied mitigation effect}. While (i) and (ii) are independent of the succession period, \textit{e.g.}, daily or weekly, our design of AQ around \textit{weekly} alternations provides a third advantage:\
(iii) It synchronizes the quarantine phase with the suspected incubation period of each cohort, hence systematically pruning out the invisible SARS-CoV-2 spreaders.
Such synchronization can readily generalize to other infections, by accordingly tuning the AQ periodicity.

Alternating quarantine can be implemented as an \textit{exit strategy}, following a period of suppression via population-wide quarantine. As such, it allows a gradual reigniting of a dormant economy, while minimizing the risk of a recurring outbreak. However, our results indicate that it can also serve as a primary mitigation strategy, with comparable impact to that of a strict population-wide quarantine (Fig.\ \ref{PWQ}). AQ should be further enforced with complementary measures, such as testing (Fig.\ \ref{Testing}) and selective protection of vulnerable populations (Supplementary Section 3).

A crucial strength of AQ is its robustness against defection, under some conditions withstanding as much as $20\%$ violators. Nevertheless, we believe that the weekly relief, allowing people an outlet to continue their activity for half of the time, may, itself, increase cooperation levels. Indeed, while a complete lock-down is extremely stressful for the individual, the AQ bi-weekly routine relaxes the burden, and may encourage compliance. Moreover, with workplaces and schools forced to operate in fully partitioned shifts, and with our suggested mobile app and logistic support network, the implementation of AQ has little dependence neither on self-motivation nor on externally enforced cooperation (Fig.\ \ref{Implementation}). Indeed, schools and employment will naturally drive the population between activity and inactivity, with \textit{enforcement} only required to treat outdoor recreation - which, in any case, has little contribution to the infection.

More broadly, we consider the fact that there is, inherently, some level of uncertainty regarding the disease parameters. We therefore examined the worst case scenario, in which the infection rate during the active weeks is the same as that of the pre-mitigation spread. In practice, however, we expect many additional measures to be implemented in parallel to the quarantines, such as extended testing for infections, face-masks and strict hygienic regulations at the workplace. At the least, we expect standard prophylactic behavior, such as avoiding contact or banning social gatherings, to be observed also during each cohort's active week. Such norms, that will continue until COVID-19 is fully eradicated, will further push down $\beta$, enhancing the effectiveness of our strategy even beyond the reported results.

Here, we have mainly discussed the epidemiological merits of AQ, and its implementation, in broad strokes, as a national strategy. In practice, different societies, as well as different economic sectors, will require specific adaptations. For example, while AQ is naturally compatible with non-professional industries, in which workers can be arbitrarily partitioned into shifts, it becomes more challenging in professional workplaces, where key personnel may be irreplaceable. Specific solutions, therefore, must be tailored to accommodate different economies and sectors. In light of AQ's unambiguous mitigating advantage, we believe such adaptations are, by far, worth the effort.

{\color{blue} \textbf{Data availability}}.
All codes to reproduce, examine and improve our proposed analysis are available at
\href{https://github.com/drormeidan/ALDCOVID19}{https://github.com/drormeidan/ALDCOVID19}.

\clearpage

\begin{Frame}
\textbf{\large \color{blue} Box I. Smooth partitioning}.\
Assigning all individuals into cohorts may seem to require coordination that is difficult to scale at a national or regional level. Here, we offer a scheme to naturally achieve a smooth partition, minimizing both economic and individual stress $\bullet$ \textbf{Employers}.\ Will be allowed to resume activity, conditional on working in fully-separated shifts. They will be given time to partition their workforce into two cohorts, $E_1$ and $E_2$, optimal for their business considerations. During this time, employers can also make other arrangements, such as training workers from $E_1$ to substitute for those in $E_2$, etc.\ $\bullet$ \textbf{Local authority}.\ Will inform all citizens of their cohort assignment, $R_1$ or $R_2$, based on, \textit{e.g}., living address. Employers will update their lists $E_1,E_2$ accordingly $\bullet$ \textbf{Conflict resolution}.\ Conflicts can arise either due to employer needs or to individual preferences. For example, if an employer detects an unbridgeable discrepancy between the official assignment of a worker ($R_1,R_2$) and their professional needs ($E_1,E_2$). Similarly, an individual may wish to switch cohort for personal reasons, \textit{e.g}., to tend for a family member in the opposite cohort. To resolve such conflicts, citizens will be given the opportunity, until a preset date $T_f$, to file for transition, \textit{of their entire household}, between $R_1$ and $R_2$. The local authority will update their lists accordingly, informing schools and other relevant institutions of the transition $\bullet$ \textbf{Flexibility}.\ The result is a friction-less scheme, essentially accommodating \textit{all} transition requests. This is enabled thanks to the fact that AQ does not require a precise $50:50$ partitioning. Hence, to allow a smooth and efficient split, both for the individual and for employers, the scheme is designed to be as flexible as possible $\bullet$ \textbf{No micro-management}.\ By the deadline $T_f$ society will naturally be divided into two cohorts, in which all employees/employers are granted their ideal work schedule. The local authorities need not micro-manage this partitioning, just track it. Once the partition is set at $T_f$, no further transitions are enabled.
\end{Frame}

\clearpage

\clearpage

\clearpage

\clearpage

\clearpage

\clearpage

\clearpage

\clearpage

\clearpage

\begin{center}
\textbf{Fig. 4: The impact of Alternating quarantine. Full size.}
\end{center}

\begin{figure}[ht]
\centering
\includegraphics[width=17cm,angle=90]{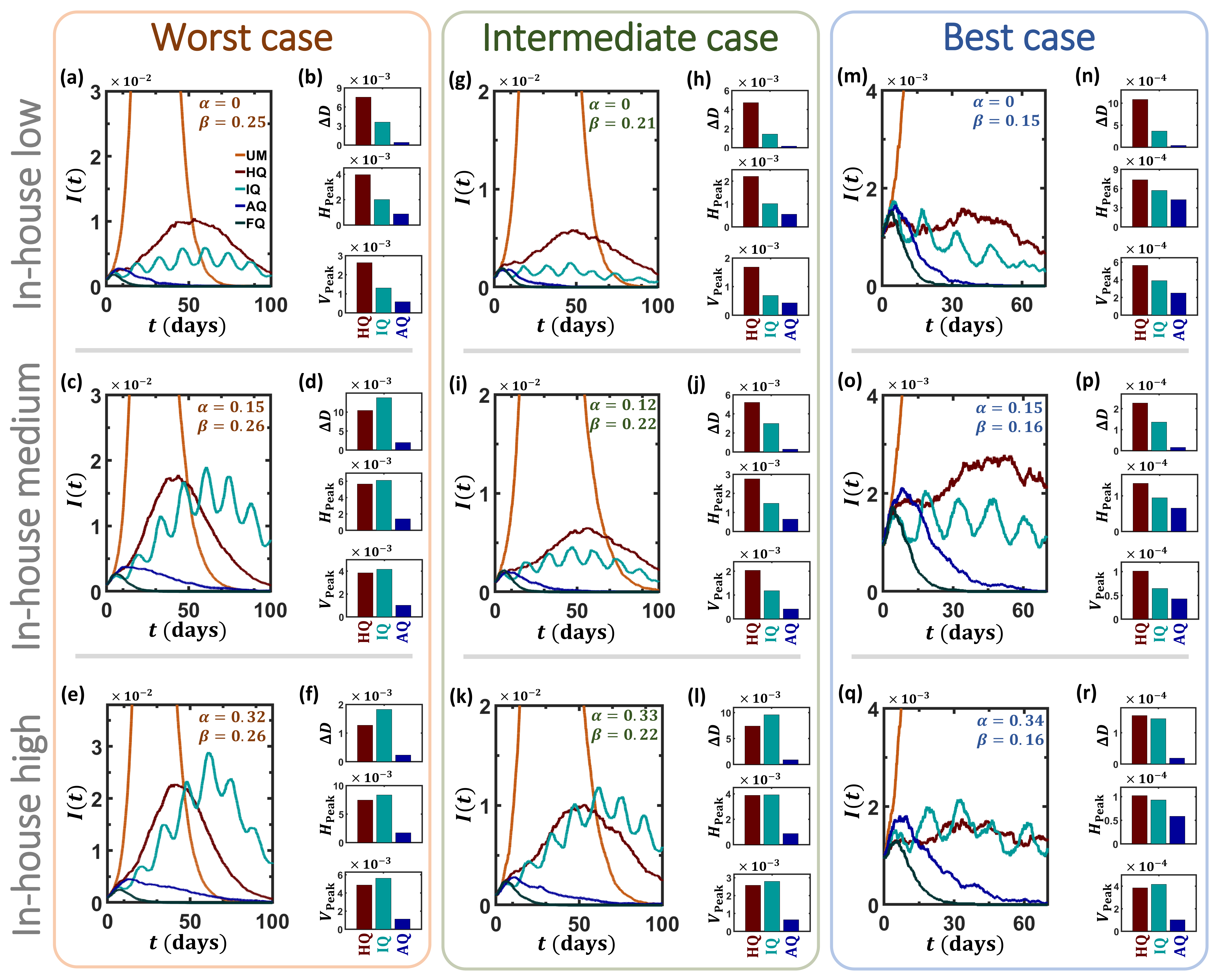}
\end{figure}

\clearpage
\bibliographystyle{unsrt}

\clearpage

\allowdisplaybreaks

\expandafter\def\expandafter\normalsize\expandafter{%
	\normalsize
	\setlength\abovedisplayskip{0pt}
	\setlength\belowdisplayskip{5pt}
	\setlength\abovedisplayshortskip{0pt}
	\setlength\belowdisplayshortskip{5pt}
}
\renewcommand{\baselinestretch}{1.1}
\setlength{\parindent}{0em}
\setlength{\parskip}{5pt}

\definecolor{Gray}{gray}{0.75}

\renewcommand{\S}{\mathbb{S}}
\newcommand{\E}{\mathbb{E}}
\newcommand{\I}{\mathbb{I}}
\newcommand{\R}{\mathbb{R}}
\newcommand{\G}{\mathbb{G}}
\newcommand{\D}{\mathbb{D}}
\newcommand{\V}{\mathbb{V}}
\newcommand{\HH}{\mathbb{H}}

\newcommand{\rL}{{\rm L}}
\newcommand{\rH}{{\rm H}}
\newcommand{\rR}{{\rm R}}
\newcommand{\rF}{{\rm F}}
\newcommand{\rD}{{\rm D}}
\newcommand{\rC}{{\rm C}}
\newcommand{\rM}{{\rm M}}
\newcommand{\rS}{{\rm S}}
\newcommand{\rV}{{\rm V}}
\newcommand{\LC}{{\rm LC}}
\newcommand{\LD}{{\rm LD}}
\newcommand{\FC}{{\rm FC}}
\newcommand{\FD}{{\rm FD}}
\newcommand{\rT}{{\rm T}}
\newcommand{\AS}{{\rm AS}}
\newcommand{\NS}{{\rm NS}}
\newcommand{\PS}{{\rm PS}}

\newcommand{\dE}[2]{\dod{\E_{#1+} #2}{t}}
\newcommand{\de}[2]{\dod{E_{#1+}^{\rm #2}}{t}}

\newcommand\infectious[1]{{\left(I^{\rm #1}_\AS(t)+I^{\rm #1}_\PS(t)\right)}}
\newcommand\infectiousalt{{\sum_{x\in\{\rm FC,D\}}\infectious{x}}}
\renewcommand\infectiousalt{{\left(I^{FC}_\AS(t)+I^{FC}_\PS(t)+I^{D}_\AS(t)+I^{D}_\PS(t)\right)}}

\newcommand{\equations}[2]
{\begin{eqnarray}
\nonumber
\\[5pt]
\dod{S^{\rm #2}}{t} &=& \ifthenelse{\equal{#2}{LC}}{0}{- \beta S(t)#1}
\label{Eq:#2:s}
\\[5pt]
\dod{E^{\rm #2}_\AS}{t} &=& \ifthenelse{\equal{#2}{LC}}{}{ \beta S(t)#1} - P_1*\de{\AS}{#2}
\label{Eq:#2:ens}
\\[5pt]
\dod{E^{\rm #2}_\rS}{t} &=& \ifthenelse{\equal{#2}{LC}}{}{(1-P_\AS)\beta S(t) #1} - P_1*\de{\rS}{#2}
\label{Eq:#2:es}
\\[5pt]
\dod{I^{\rm #2}_\AS}{t} &=& P_1*\de{\AS}{#2} - P_2*\de{\AS}{#2}
\label{Eq:#2:Ins}
\\[5pt]
\dod{I^{\rm #2}_\PS}{t} &=& P_3*\de{\rS}{#2} - P_4*\de{\rS}{#2}
\label{Eq:#2:Ips}
\\[5pt]
\dod{I^{\rm #2}_\rM}{t} &=& P_\rM P_4*\de{\rS}{#2} - r_{\rM\rR}I^{\rm #2}_\rM
\label{Eq:#2:Im}
\\[5pt]
\dod{I^{\rm #2}_\rS}{t} &=& P_\rS P_4*\de{\rS}{#2} - r_{\rS\rH}I^{\rm #2}_\rS
\label{Eq:#2:Is}
\\[5pt]
\dod{I^{\rm #2}_\rC}{t} &=& P_\rC P_4*\de{\rS}{#2} - r_{\rC\rV}I^{\rm #2}_\rC
\label{Eq:#2:Ic}
\\[5pt]
\dod{H^{\rm #2}}{t} &=& r_{\rS\rH}I^{\rm #2}_\rS - P_{\rH\rR}r_{\rH\rR}H^{\rm #2}-P_{\rH\rD}r_{\rH\rD}H^{\rm #2}
\label{Eq:#2:H}
\\[5pt]
\dod{V^{\rm #2}}{t} &=& r_{\rC\rV}I^{\rm #2}_\rC - P_{\rV\rR^{\rm #2}}r_{\rV\rR}V-P_{\rV\rD}r_{\rV\rD}V^{\rm #2}
\label{Eq:#2:V}
\\[5pt]
\dod{R^{\rm #2}}{t} &=& P_2*\de{\AS}{} + P_{\rH\rR}r_{\rH\rR}H + P_{\rV\rR}r_{\rV\rR}V+ r_{\rM\rR}I_\rM
\label{Eq:#2:R}
\\[5pt]
\dod{D^{\rm #2}}{t} &=& P_{\rH\rD}r_{\rH\rD}H^{\rm #2}+P_{\rV\rD}r_{\rV\rD}V^{\rm #2}
\label{Eq:#2:D}
\end{eqnarray} }

\newcommand{\equationssingle}{\equations{\infectious{}}{}}
\newcommand{\equationslockdown}{\equations{}{LC}}
\newcommand{\equationsfree}{\equations{\infectiousalt}{FC}}
\newcommand{\equationsdefectors}{\equations{\infectiousalt}{D}}

\begin{center}
{\Huge \color{blue} \textbf{Alternating quarantine for \\ \vspace{5mm} sustainable epidemic mitigation}}

\vspace{15mm}

{\huge \color{blue} \textbf{Supplementary Information}}
\end{center}

\vspace{10mm}


\clearpage


\section{Modeling the spread}
\label{section: model description}

\subsection{The unmitigated spread}

\subsubsection{Population network}

We consider $M$ households $h = 1,\dots,M$, each including $m_h$ individuals, where $m_h$ is a random variable extracted from the household size distribution $P(m)$ (Table \ref{table: household data by country}). Together these households comprise the total population, in which the number of individuals $i = 1,\dots,N$ is given by $N = \sum_{h = 1}^{M} m_h$. Hence, the $i$-th individual, resides together with her $m_h - 1$ cohabitants at her household $h(i)$. Taking $M = 4 \times 10^3$, and the average household size to be $\av m = 2.5$, we arrive at a total population of $N = 10^4$ individuals.

To construct the social network $\m Gij$, we consider two types of links:\ Within a household there are no barriers, hence the in-house connection network $B_{ij}$ is simply a union of disjoint cliques representing $(i,j) \in B_{ij} \Longleftrightarrow h(i) = h(j)$. This results in $M$ isolated cliques whose size is distributed via $P(m)$. Out of home connections, $A_{ij}$, can be potentially drawn between any pair of nodes $i,j$, capturing external social links, occurring at work, school or other public places. The external network $\m Aij$ can be admit any desired degree distribution $P(k)$ via the configuration model framework \cite{Newman2010}. In our simulations we used two archetypal constructions - the Erd\H{o}s-R\'{e}nyi (ER) random graph (main text), in which $P(k)$ is bounded, and a scale-free (SF) network, where $P(k) \sim k^{-3}$ (Sec.\ \ref{section: scale-free}). In both cases we set the average degree to $\av k = 15$. The final network $\m Gij$ contains all links in $\m Aij$ and $\m Bij$.

\subsubsection{Temporality}

The links in $\m Gij$ are not constantly active. Rather, they represent \textit{potential} infectious interactions, switching between periods of activity, when infections can take place, and inactivity, when infections are barred. Throughout the daily cycle we have $\m Aij$ active during the day, 8:00 AM to 8:00 PM, and $\m Bij$ active during the after-hours, 8:00 PM to 8:00 AM the next day. This captures a typical routine, in which individuals interact sporadically, \textit{i.e}.\ links are switched on and off, out of home in the day-time, and in-house at night.

Dividing each day into $15$ minute segments, $\Delta t$, we generate a random sequence of temporal activity/inactivity instances for $\m Aij$ and $\m Bij$. During the day, the probability for activation of each link in $\m Aij$ per interval $\Delta t$ is set to $p_A$. Similarly, during the night we have probability $p_B$ for $\m Bij$ activation. The result is a stochastic pattern of potentially infectious interactions, in which the idling time between subsequent instances of activity follows a geometric distribution $\text{Geom}(p_A)$ or $\text{Geom}(p_B)$ for $\m Aij$ and $\m Bij$, respectively. On average, we have $\m Aij$ links active

\begin{equation}
T_1 = 12p_A
\label{Tout}
\end{equation}

hours, and $\m Bij$ active for

\begin{equation}
T_2 = 12p_B
\label{Tin}
\end{equation}

hours, per daily cycle.

There are two exceptions to the above random activation rules:\
\begin{itemize}
\item
\textbf{Isolation}.\ In case node $i$ is known to be infected, \textit{i.e}.\ symptomatic (Sec.\ \ref{section: dynamics}), then $i$'s entire household $h(i)$ enters isolation. All nodes in $h(i)$ remain at home until the household is cleared to retain its activity. Under these conditions only $B_{ij}$ is activated with probability $p_B$ throughout the entire $24$ hour cycle, and $\m Aij$ links remain idle. Consequently, when in isolation, in-house interactions become more extensive, as they have more potential instances of infection, then during periods of normal activity.
\item
\textbf{Collocation}.\ For consistency, if at a certain instance, both links $(i,j)$ and $(i,k)$ are simultaneity active, then the triad link $(j,k)$ is also activated. Indeed, a concurrent collocation of $i,j$ and $i,k$, implies, by transitivity, an inevitable collocation also of $j,k$. This alllows us to capture potential correlations in the temporal patterns of the interactions.
\end{itemize}

A summary of all temporal network parameters appears in Table \ref{table: network parameters}.

\setlength{\heavyrulewidth}{0.08em}
\setlength{\lightrulewidth}{0.05em}
\setlength{\cmidrulewidth}{0.03em}

\begin{table}[ht]
  \centering
\begin{tabular}{lll}
  \toprule
  \hline
\textbf{Parameter} & \textbf{Description} & \textbf{Value} \\
\midrule
$N$ & Population size  & $10^4$ \\
$M$ & Number of households  & $4 \times 10^3$ \\
$p(k)$ & $\m Aij$ degree distribution  & Erd\H{o}s-R\'{e}nyi or scale-free \\
$p(m)$ & Household size distribution  & Empirically obtained, Table \ref{table: household data by country} \\
$\av k$ & Average $\m Aij$ degree & $15$ \\
$\av m$ & Average household size & $2.5$ \\
$p_A$ & Probability of $\m Aij$ link activation & Varied \\
$p_B$ & Probability of $\m Bij$ link activation & Varied \\
$T_1$ & Mean daily infection time via $\m Aij$ & Eq.\ (\ref{Tout}) \\
$T_2$ & Mean daily infection time via $\m Bij$ & Eq.\ (\ref{Tin}) \\
$\alpha$ & Fraction of in-house infections & Extracted from data/simulation \\
$\beta$ & infection growth rate & Extracted from data/simulation \\

  \bottomrule
\end{tabular}  \caption{\textbf{The parameters governing the temporal network}.\ We list the relevant quantities underlying our temporal network framework. Parameters that are unknown are varied to capture the breadth of different epidemiological scenarios. For example $P(k)$ is set to be both bounded (main text) or scale-free (Sec.\ \ref{section: scale-free}); $p_A, p_B$ are varied in our simulations. Parameters $\alpha,\beta$ are not set, but rather extracted from the observed spread, as explained in Sec.\ \ref{section: alphabeta}.}
  \label{table: network parameters}
\end{table}

\subsection{Mitigation}

During mitigation, the quarantined households express only $\m Bij$ throughout the $24$ hour cycle, with all their $\m Aij$ links rendered inactive. If a household member is defective, their $\m Aij$ links continue to activate as usual. Partitioning the population, as in AQ or HQ, for example, is done at household level - namely households are randomly split among the cohorts. In each realization, we instigate the mitigation at a time $t_0$ when the fraction of infections $I(t = t_0)$ exceeds a \textit{significant} threshold. We set this threshold at

\begin{equation}
I(t = t_0) = \dfrac{\ln N}{N},
\label{t0}
\end{equation}

namely the time point where the total infected population is of the order of $\ln N$.

\subsection{Disease dynamics}
\label{section: dynamics}

\begin{figure}
\centering
\includegraphics[width=\textwidth]{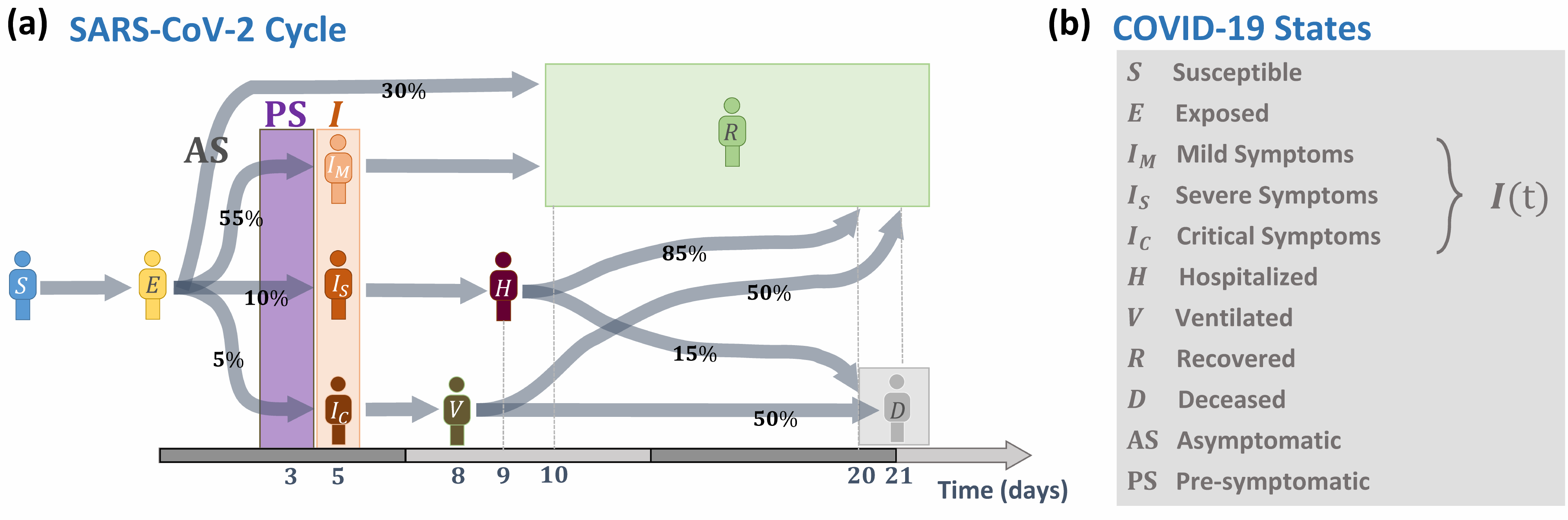}
\caption{\textbf{The infection cycle of SARS-CoV-2}. Extracted from Fig.\ 1 of the main text.}
\label{DiseaseCycle}
\end{figure}

We begin with a fully susceptible ($\S$) population, and introduce a small fraction of exposed ($\E$) individuals. The potential transitions that ensue are shown in Fig.\ \ref{DiseaseCycle}, whose main transitions include:\

\begin{itemize}
\item
\textbf{Infection}.\ At any encounter between a susceptible node $i$ and a pre-symptomatic or infected node $j$, $i$ may become exposed. By \textit{encounter} we relate to an instance $\Delta t$ in which the $i,j$ link in $\m Aij$ \textit{or} $\m Bij$ is active. The probability of infection at each encounter depends on the nature of the link, set to $p_1$ for $\m Aij$ links and $p_2$ for $\m Bij$. External interactions $\m Aij$, between associates, are typically less physical than in-house interactions between \textit{e.g.}, family members, hence, typically $p_2 > p_1$. In practice, however, we can incorporate these probabilities into the encounter probabilities themselves, $p_A, p_B$. Indeed, stating that $i$ and $j$ interact with probability $p_A$ and then infect with probability $p_1$, is equivalent to setting their interaction probability to $p_Ap_1$, and having infections occurring with $100\%$ certainty. Hence, for simplicity we set $p_1 = p_2 = 1$, and encapsulate the infection probabilities within the parameters $T_1,T_2$ in (\ref{Tout}) and (\ref{Tin}).

\item
\textbf{Infection classification}.\ During the simulation of the spread we keep count of the type of each infection. Infections occurring via $\m Bij$ links add to the in-house infection count $\theta_{\text{In}}$; infections occurring out of home, through $\m Aij$ contribute to $\theta_{\text{Out}}$.

\item
\textbf{Infection cycle}.\ Once a node becomes exposed it begins to transition between states according to Fig.\ \ref{DiseaseCycle}a. Exposed nodes have contracted the virus, but are not yet infectious. These nodes are randomly split between $\E_\rS$ with probability $p_\rS$ and $\E_\AS$ with probability $p_\AS = 1 - p_\rS$. This decides whether these nodes are \emph{pre-symptomatic}, and eventually \textit{will} develop symptoms, or \emph{asymptomatic}, reaching recovery $\R$ without even experiencing symptoms. The remaining disease cycle continues according to the illustration. For example, nodes in $\E_\rS$ will later transition to one of the infected stated $\I_\rM, \I_\rS$ or $\I_\rC$ with probabilities $p_\rM = 0.55, p_\rS = 0.1$ and $p_\rC = 0.05$, respectively; the remaining $30\%$ are accounted for in the $\E_\AS$ trajectory. Similarly, $\I_\rS$ nodes, after some time enter the hospitalized state $\HH$, after which the recover with probability $p_\mathrm{HR} = 0.85$, and decease with probability $p_\mathrm{HD} = 0.15$.

Note that ventilated individuals $\V$ are, by definition, also hospitalized. However, in out implementation we consider these as two isolated groups, \textit{i.e}.\ ventilated vs.\ hospitalized without ventilation. Therefore, at all times we have $\S(t) + \E(t) + \I_\NS(t) + \I(t) + \R(t)+ \V(t) + \HH(t) + \D(t) = N$, comprising the entire population. Presenting our results we used the normalized compartments $S(t) = \S(t)/N, E(t) = \E(t)/N, I_\NS(t) = \I_\NS(t)/N, \dots$, which satisfy

\begin{equation}
S(t) + E(t) + I_\NS(t) + I(t) + R(t)+ V(t) + H(t) + D(t) = 1.
\label{Normalization}
\end{equation}

\item
\textbf{Transition times}.\ The amount of time a node remains at a state $\mathbb{X}$ (other than $\S$) is chosen at random according to probability density $P_\mathbb{X}(t)$. We identify specific processes for which  variability in the transition time may impact the effectiveness of alternating quarantine (AQ). For example, the time from exposure to infectiousness, or the time for asymptomatic individuals to recover are crucial. Deviations from the mean in these transition times may interfere with AQ's disease cycle synchronization. For instance, if a node remains asymptomatic for, \textit{e.g.}, $3$ weeks, which is beyond the average time to recovery, its infectiousness may spillover between AQ's subsequent activity cycles, allowing it to resume activity while still infectious. Similarly, if the pre-symptomatic stage is extended significantly beyond the $5$ day average, an infected individual in week $1$ may not yet develop symptoms during their isolation at week $2$, once again, reducing the efficiency of AQ in removing \textit{invisible spreaders}. Therefore, for these highly relevant transitions we placed a special emphasis to avoid underestimating their potential time-scale heterogeneity. In particular we identified four relevant processes:\ $P_{\E_\AS}(t)$, the time until an exposed asymptomatic individual ($\E_\AS$) becomes infectious ($\I_\AS$); $P_{\I_\AS}$, the time until an asymptomatic infectious node recovers; $P_{\E_\rS}(t)$, the time until an exposed pre-symptomatic individual ($\E_\rS$) becomes infectious ($\I_\PS$) and $P_{\I_\rS}(t)$, the time for an infectious pre-symptomatic ($\I_\PS$) to show symptoms ($\I_\rM,\, \I_\rS$ or $\I_\rC$). For these four functions we used a Weibull distribution, as explained in Sec.\ \ref{subsection: estimating Weibull}. This distribution allows us to capture the potentially variable time-scales across the population, thus testing AQ under realistically challenging conditions.

\end{itemize}

\subsection{Evaluating \texorpdfstring{$\alpha$}{alpha} and \texorpdfstring{$\beta$}{beta}}
\label{section: alphabeta}

Our model parameters control the spreading dynamics via the temporal probabilities $p_A$ and $p_B$, and their subsequent $T_1$ and $T_2$ in (\ref{Tout}) and (\ref{Tin}), that govern the rate of infections in and out of home. Once these parameters are set, the simulating results of the unmitigated spread allow us to evaluate the infection growth rate $\beta$ and the in-house transmission rate $\alpha$:\

\begin{itemize}
\item
\textbf{In-house infection rate $\alpha$}.\
During the simulation we keep count of the source of all infections. Infections occurring via $\m Aij$, amounting to $\theta_{\rm Out}$ are external, while those that transmit along $\m Bij$ links, $\theta_{\rm In}$, are internal. The parameter $\alpha$ captures the percentage of transmissions that occurred in-house as

\begin{equation}
\alpha = \dfrac{\theta_{\rm In}}{\theta_{\rm In} + \theta_{\rm Out}}.
\label{supp eq: Alpha}
\end{equation}

\item
\textbf{The infection growth rate $\beta$}.\
To evaluate $\beta$ we observe the overall infected population $I(t)$ vs $t$ at the early stages of the spread, and fit it to an exponential of the form

\begin{equation}
I(t)\sim e^{\beta t}.
\label{supp eq: Beta}
\end{equation}

Obtaining the slope of the resulting growth on semi-logarithmic axes we extract $\beta$ from the simulation results. Note that $\beta$ depends on the \textit{slope}, not on the pre-factor, hence it is insensitive to the size of the initial outbreak, or to the fraction of cases detected via testing, providing a fair comparison between different realizations or empirical datasets.

A crucial point is to select the range in $t$ from which to extract the slope. Indeed, for very small $t$, due to the stochastic nature of our simulations, $\I(t)$ is still small, and still discrete. In this limit, the observed results are subject to high levels of noise and may not yet exhibit a clear exponential behavior. On the other hand if $t$ is too large, we approach the peak of $I(t)$, where the exponential approximation fails again, this time due to the accumulation of herd immunity. Therefore, to be consistent across all our simulations we evaluated $\beta$ from the time window

\begin{equation}
\frac{t^{\star}}{4} \le t \le \frac{t^{\star}}{2}
\label{TimeWindow}
\end{equation}

where $t^{\star} = {\rm argmax} I(t)$ is the time of peak infection. Evaluating $\beta$ from empirical data is explained in Sec. \ref{section: parameter estimation}.

\end{itemize}

Note, that $\alpha$ and $\beta$ are \textit{not} the model parameters. Rather they emerge from the stochastic simulation results, after setting the model parameters $p_A$ and $p_B$. Therefore, we do not have direct control over these parameters, as seen in, \textit{e.g.}, Fig. 4 of the main text, where $\alpha,\beta$ were only approximately equal across the different panels. Roughly speaking, we can link these parameters to each other. A large $p_A,p_B$, for example enhances transmission, and hence increases $\beta$. The parameter $\alpha$, on the other hand, grows as $p_B$ is increased and $p_A$ is decreased, capturing a state in which in-house transmissions are more prevalent than external ones.

\section{Results obtained under a scale-free \texorpdfstring{$\m Aij$}{Aij}}
\label{section: scale-free}

\begin{figure}
\centering
\includegraphics[width=\textwidth]{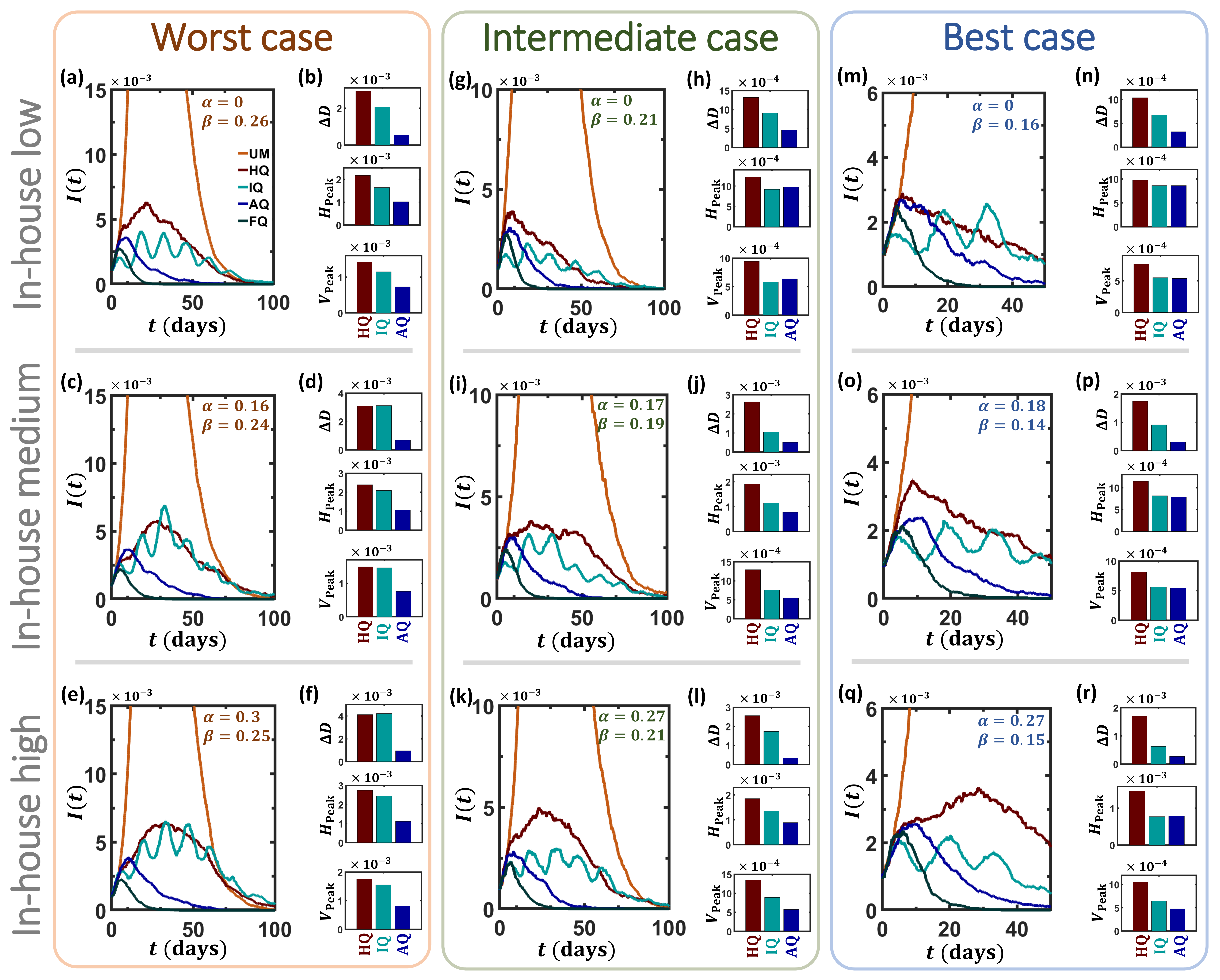}
\caption{\textbf{The impact of alternating quarantine for a scale-free $\m Aij$.} Reconstructing Fig.\ 4 of the main text, this time using a scale-free external network.}
\label{fig: Result_SF}
\end{figure}

\begin{figure}
\centering
\includegraphics[width=\textwidth]{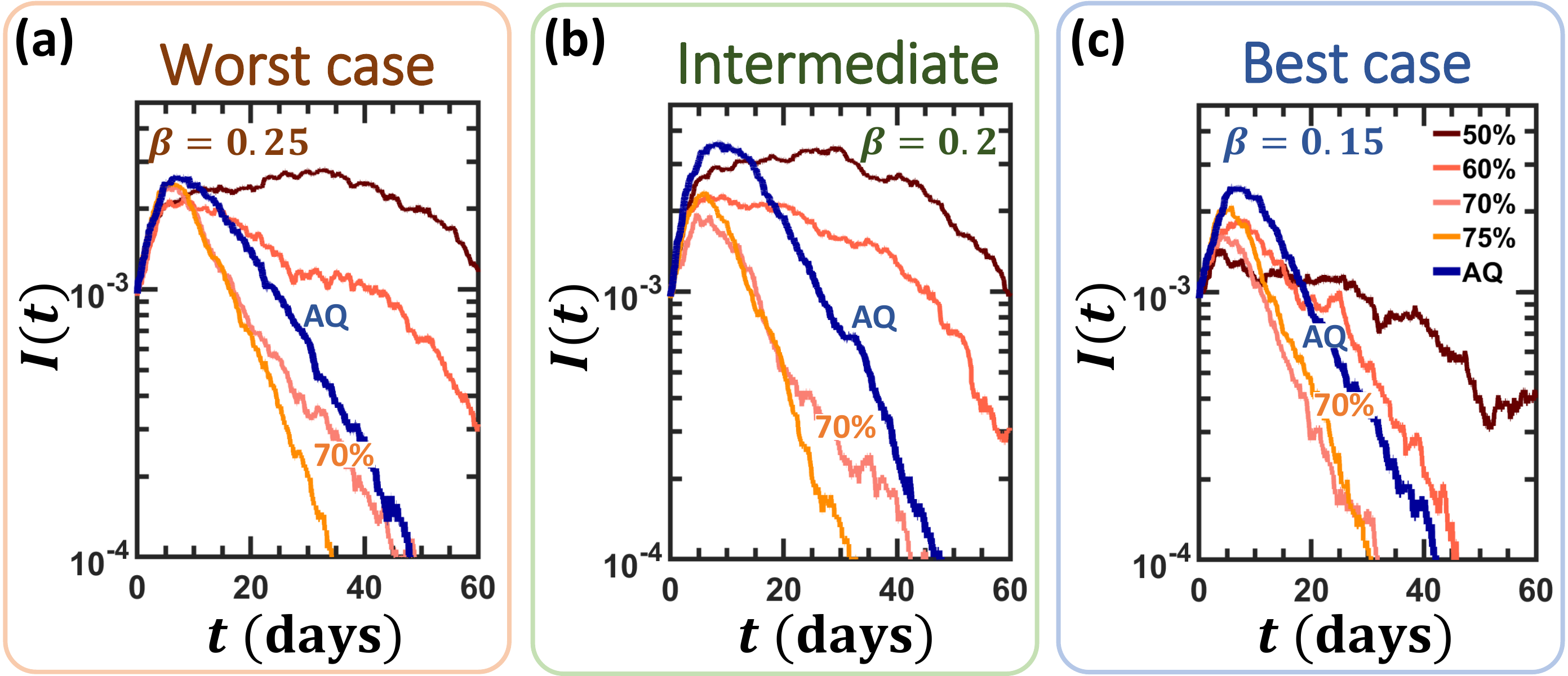}
\caption{\textbf{Alternating quarantine vs. population-wide quarantine on a scale-free $\m Aij$.} Reconstructing Fig.\ 7 of the main text, this time using a scale-free external network.}
\label{fig: PWQ_SF}
\end{figure}

Our strategy in testing AQ is to examine it systematically under varying relevant scenarios. Specifically, for unknown parameters, such as $\alpha$ and $\beta$, we simulated an array of different setting, scanning the space of potential $\alpha,\beta$ values (\textit{e.g}., Fig.\ 4 of main text). Other unknown factors relate to the structural characteristics of the external network $\m Aij$. Most importantly, in the context of epidemic spreading - its degree distribution, which has been shown to significantly impact the patterns of spread \cite{PastorSatorras2015}. To eliminate this potentially confounding factor we now re-examine AQ, repeating our simulations, this time extracting $\m Aij$ from the scale-free network ensemble ($P(k) \sim k^{-3}, N = 10^4, \av k = 15$). We find, in Figs.\ \ref{fig: Result_SF} and \ref{fig: PWQ_SF} that AQ continues to provide the optimal mitigation also under these conditions.

\section{Alternating quarantine under selective isolation}
\label{AQ selective isolation}

\begin{figure}
\centering
\includegraphics[width=\textwidth]{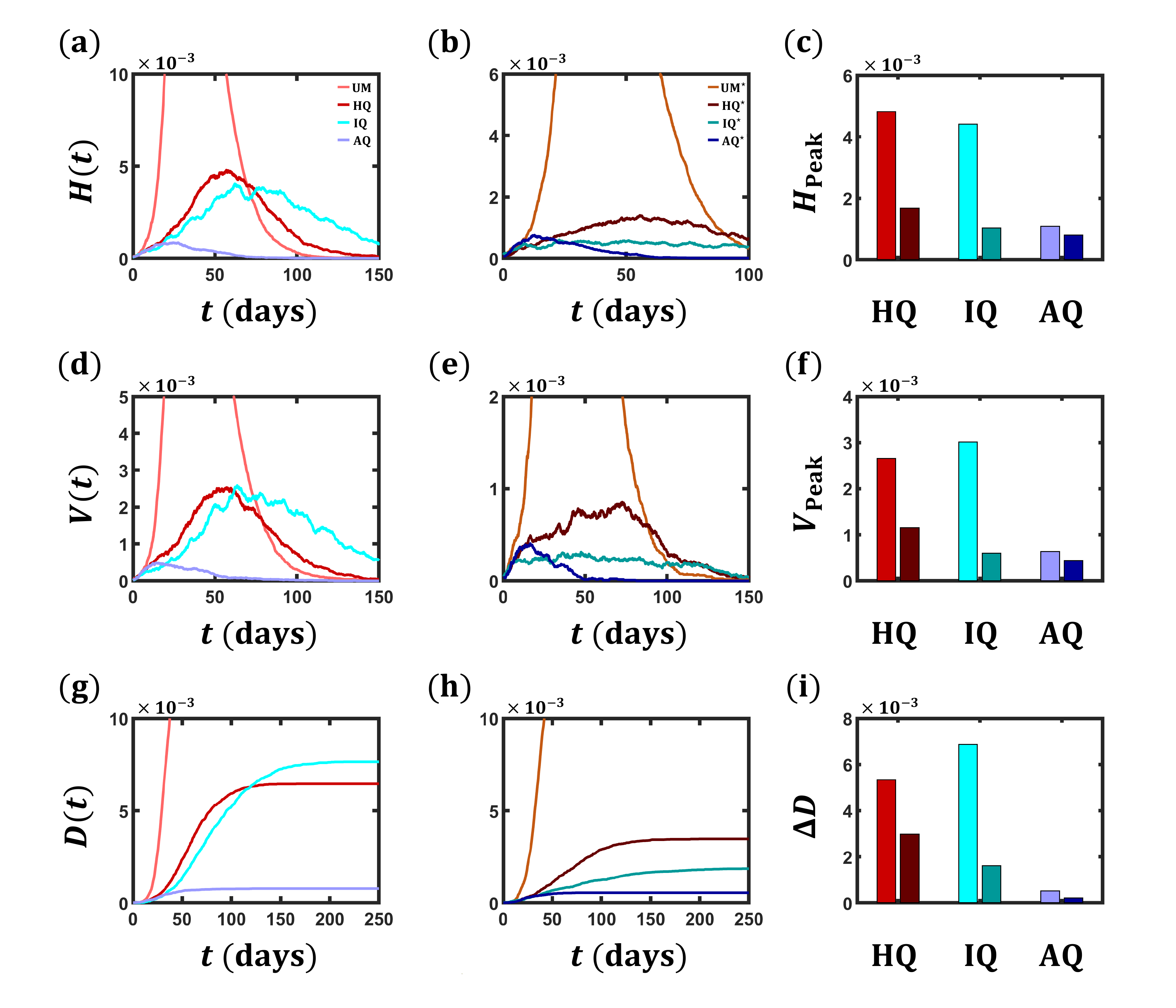}
\caption{\textbf{The impact of selective isolation on the different mitigation approaches}. (a) The fraction of hospitalized individuals $H(t)$ vs.\ $t$ under no mitigation (light orange, UM), intermittent quarantine (IQ, light turquoise), half quarantine (HQ, light red) and alternating quarantine (AQ, light blue). (b) Similar results (dark colors), this time with selectively isolating the vulnerable population. (c) Peak hospitalization under all strategies with (light) and without (dark) selective isolation.
(d) - (f) Ventilated population $V(t)$ vs. $t$ with/without selective isolation.
(g) - (i) Mortality $D(t)$ and the residual mortality $\Delta D$.}
\label{fig: spi}
\end{figure}

In the main take we used a \textit{typical} disease cycle, capturing the average individual's response to SARS-CoV-2. We now consider two parallel cycles, one for healthy individuals and the other for the vulnerable, such as people with background diseases or the elderly. The two cycles differ mainly in their transition probabilities. For example, while only $4\%$ of the healthy individuals develop critical symptoms ($\I_\rC$), among the vulnerable population the number is set to $10\%$. The complete disease cycle for the Typical, Healthy and Vulnerable population \cite{si: Mallapaty2020,si: Bonafe2020,si: Xie2020} appears in Table \ref{table: probabilities}.

To track the spread in the presence of healthy/vulnerable populations we repeated the simulation described in Sec.\ \ref{section: model description}, this time splitting the population into $80\%$ healthy and $20\%$ vulnerable nodes \cite{UNPopulation}. We track three indicators that help us assess the performance of all strategies (Fig.\ \ref{fig: spi}):\ Hospitalization rate $H(t)$, ventilation rate $V(t)$ and mortality $D(t)$. As expected, AQ (Fig.\ \ref{fig: spi})a,d,g, light blue) continues to outperform IQ (light turquoise) and HQ (light red) also under this variable disease cycle.

Next, we added an additional component of selective isolation, in which the vulnerable nodes ($20\%$) remain under constant quarantine. For example, in AQ this implies that the weekly alternations are limited only to the healthy $80\%$. Under these conditions the vulnerable individuals cannot be infected via external links $\m Aij$. They can still, however, experience secondary infection through $\m Bij$, in case one of their healthy cohabitants contracted the virus. As expected, such selective isolation enhances the performance of all the strategies, lowering hospitalization, ventilation and mortality (Fig.\ \ref{fig: spi}b,e,h). This improvement, we emphasize, is not unique to AQ, making it clear that selective isolation is a desirable component within any mitigation strategy. In Fig.\ \ref{fig: spi}c,f,i we present our three performance measures, $H_{Peak}$, $V_{Peak}$ and $\Delta D$ with (dark) and without (light) selective isolation, further indicating the importance of protecting the vulnerable population.

\setlength{\heavyrulewidth}{0.08em}
\setlength{\lightrulewidth}{0.05em}
\setlength{\cmidrulewidth}{0.03em}

\begin{table}[ht]
  \centering
\begin{tabular}{lllll}
  \toprule
  \hline
\textbf{Probability} & \textbf{Typical} & \textbf{Healthy} & \textbf{Vulnerable} \\
\midrule
$p_{AS}$ & 0.3  & 0.32 & 0.25 \\
$p_{M}$ & 0.55  & 0.56 & 0.45 \\
$p_{S}$ & 0.1  & 0.08 & 0.2 \\
$p_{C}$ & 0.05  & 0.04 & 0.1 \\
$p_{HR}$ & 0.85  & 0.86 & 0.79 \\
$p_{HD}$ & 0.15  & 0.14 & 0.21 \\
$p_{VR}$ & 0.5  & 0.5 & 0.5 \\
$p_{VD}$ & 0.5  & 0.5 & 0.5 \\

  \bottomrule
\end{tabular}  \caption{\textbf{Transition probabilities between COVID-19 states.} We constructed three disease cycles - the Typical cycle, used in the main text, the Healthy cycle, capturing the impact of the disease on healthy individuals, and the Vulnerable cycle, adapted to individuals of age or ones with pre-existing conditions.}
  \label{table: probabilities}
\end{table}

\section{Data analysis and parameter selection}\label{section: parameter estimation}
\label{parameter}
\subsection{Constructing the distributions \texorpdfstring{$P_\mathbb{X}(t)$}{PXt}}
\label{subsection: estimating Weibull}

Most of the parameters described in Section \ref{section: model description} were chosen based on observed values of the characteristic SARS-CoV-2 infection cycle. For the density functions $P_\mathbb{X}(t)$, we have used a Weibull or a Geometric distribution, the former - inspired both by other infections of the Corona variety \cite{si: bar2020sars}, as well as recent inidcations pretainig to SARS-CoV-2 \cite{si: linton2020incubation,si: Lauer2020,si: backer2020incubation}. The Weibull distribution allows for potentially high variability across the population, providing a \textit{challenging} testing ground for AQ.

To estimate the parameters of the Weibull distributions we collected data on the average $T_{\rm Av}$ and median $T_{\rm Med}$ of the relevant transition times \cite{si: linton2020incubation,si: bar2020sars}. This allowed us to infer the Weibull parameters $\lambda$ and $k$ via

\begin{align*}
T_{\rm Av} &= \lambda \Gamma(1+1/k); \\
T_{\rm Med} &= \lambda (\ln 2)^{1/k}.
\end{align*}

As median values were available only for $P_{\I_\AS}$ and $P_{\I_\PS}$, we first calculated the parameter $k$ for these transitions, obtaining, for both $k=1.47$. This is not surprising as $k$, the \textit{shape} parameter, controls the type of the Weibull distribution, which is expected to be similar for processes driven by similar mechanisms. This is as opposed to $\lambda$, the \textit{location} parameter, which is not intrinsic to the shape of the distribution, but rather shifts \textit{right} or \textit{left} as the mean is changed. Hence, it is expected that $k$ is uniform for the different transition-time distributions, while $\lambda$ may change according to their mean. With this in mind, we estimated $k = 1.47$ for the other two distributions, $P_{\E_\AS}$ and $P_{\E_\rS}$, where the median was inaccessible from data. See Table \ref{table: Weibull parameters} for the different values of mean, and median we have used, and the inferred $\lambda$ and $k$.

\begin{table}[ht]
  \centering
\begin{tabular}{llllll}
  \toprule
\textbf{Duration} & \textbf{Distribution} &\multicolumn{2}{c}{\textbf{Observations}} & \multicolumn{2}{c}{\textbf{Parameters}} \\
  \cmidrule(r){3-6}
 & & \textbf{Mean} & \textbf{Median} & $\lambda$ & $k$ \\
\midrule
$P_{\E_\AS}(t)$ & Weibull & 4  & $3.44^*$ & 4.42 & $1.47^*$ \\
$P_{\I_\AS}(t)$ & Weibull & 10 & 8.6  & 11.04 & 1.47 \\
$P_{\E_\rS}(t)$ & Weibull & 2  & $1.72^*$ & 2.21 & $1.47^*$ \\
$P_{\I_\PS}(t)$ & Weibull & 5  & 4.3  & 5.52 & 1.47 \\
\midrule
 & & \textbf{Mean} & & $p$ & \\
\cmidrule(r){3-6}
$P_{\I_\rM}(t)$ & Geometric & 5  &   & 0.2 &  \\
$P_{\I_\rS}(t)$ & Geometric & 4  &   & 0.25 &  \\
$P_{\I_\rC}(t)$ & Geometric & 3  &   & 0.333 &  \\
$P_{\I_\HH}(t)$ & Geometric & 11  &   & 0.091 &  \\
$P_{\I_\V}(t)$ & Geometric & 13  &   & 0.077 &  \\
  \bottomrule
\end{tabular}  \caption{\textbf{Estimating the distribution parameters}.\ With data on the mean and median transition times, we reconstructed the distributions $P_\mathbb{X}(t)$. For the first four transitions we used a Weibull distribution, since the potentially high variability is key for testing AQ. The remaining distributions were taken to be Geometric, since only the mean matters for these transitions. Asterisked median values are reconstructed.}
  \label{table: Weibull parameters}
\end{table}

\subsection{Estimating the infection growth \texorpdfstring{$\beta$}{beta}}

As defined above, the parameter $\beta$ represents the exponential growth rate of the known infectious nodes $I(t) := I_M(t) + I_S(t) + I_C(t)$. This parameter is difficult to predict directly from the knwon disease time-scales, especially as the infection rate is hidden, hence we must infer it from observation. Moreover, as the disease progresses, precautions like social distancing and wearing masks affect both the rate of interaction and the probability of infection, leading $\beta$ to change over time. Therefore, to asses $\beta$ for the unmitigated spread, we have focus on the period \emph{before} such measures were taken.

We collected data on the number of confirmed cases in $12$ countries. These countries have been selected for their prominent number of casesm and to obtain a balanced representation between southern and northern hemisphere destinations. The data set was compiled by and obtained from the Johns Hopkins University Center for Systems Science and Engineering (JHU CSSE) on April 11\textsuperscript{th} 2020 and is available online here: \href{https://data.humdata.org/dataset/novel-coronavirus-2019-ncov-cases}{\color{blue} \textit{https://data.humdata.org/dataset/novel-coronavirus-2019-ncov-cases}} \cite{si: JohnHopkins2020}.

\begin{figure}
\centering
\includegraphics[width=0.4\textwidth]{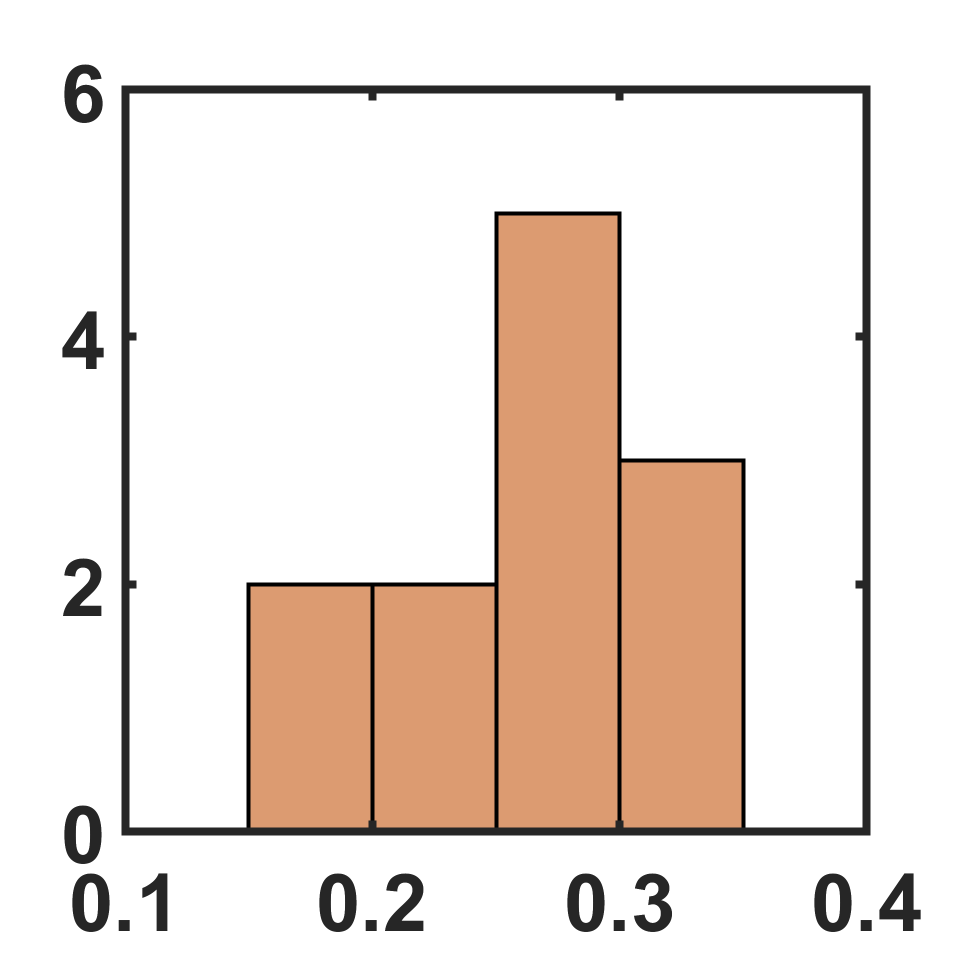}
\caption{\textbf{The variability of infection rates across countries}.\ Histogram of the estimator $\hat\beta$ values by countries.}
\label{fig: beta dist}
\end{figure}

To capture the relevant time-window for the exponential growth approximation we used data-points starting $5$ days before lock-down and ending $3$ days after it. Indeed, earlier than this point, cases may be underestimated by a yet unprepared system, and beyond this window, the lock-down may begin affecting the observed slope. As clearly seen in Fig.\ 3 of the main text, within this time-window the spread $I(t)$ can be well-approximated by an exponential growth of the form (\ref{supp eq: Beta}). To extract the slope we used linear regression on $\ln I(t)$, yielding the estimator $\hat \beta$ for the growth rate in each country, as detailed in Table \ref{table: countries data} and in Fig.\ \ref{fig: beta dist}. We find that estimators are narrowly distributed around an average of $\beta = 0.26$, the value we used as our \textit{default}, \textit{i.e}.\ unmitigated spreading parameter.

\begin{table}[ht]
  \centering
\begin{tabular}{p{3cm}p{2.2cm}p{2.2cm}p{2.2cm}p{1.7cm}p{2.8cm}}
  \toprule
\hline
\textbf{Country}	&	\textbf{Population} & \textbf{First case} & \textbf{Lock-down} &	$\hat{\beta}$\\
\midrule
Italy	&	60	&	10	&	38 &	0.32 \\
USA	&	328	&	3		&	61 &	0.3 \\
Spain	&	47	&	19	&	43 &	0.34 \\
Israel	&	9	&	36		&	54 &	0.19 \\
Germany	&	83	&	7	&	52	&	0.26 \\
Norway	&	5.4	&	38	&	49 &	0.32 \\
Colombia	&	52	&	49	&	56 &	0.28 \\
Argentina	&	45	&	45		&	50 &	0.22 \\
Netherlands	&	17	&	39		&	54 &	0.21 \\
N.\ S.\ Wales	&	8	&	5		&	52 &	0.18\\
Austria	&	9	&	35	&	45 &	0.3\\
UK & 56 & 10 & 46  & 0.29 \\
  \bottomrule
\end{tabular}  \caption{\textbf{Estimating $\beta$ per country}.\ Population is given in millions.  First case and Lock-down are given in days relative to $22/1$. The parameter $\hat{\beta}$ represents the estimation for $\beta$, as extracted from the relevant country data. See Fig.\ \ref{fig: beta dist} for a histogram of $\hat\beta$.}
  \label{table: countries data}
\end{table}

\subsection{Estimating the household size distribution \texorpdfstring{$P(m)$}{Pm}}
\label{subsection: parameter household}

We used a United Nations database \cite{si: UNHouseholds} to collect data on the distribution of household sizes across different countries. The data, summarized in Table \ref{table: household data by country}, was compiled by and obtained from the United Nations, Department of Economic and Social Affairs, Population division.

\begin{table}[ht]
  \centering
\begin{tabular}{p{3cm}p{2.2cm}p{2.2cm}p{2.2cm}p{1.7cm}p{2.8cm}}
  \toprule
\hline
\textbf{Country}	&	1 & 2-3 & 4-5 & 6 & Average \\
\midrule

Italy	&	0.31	&	0.47	&	0.21 &	0.01 & 2.4  \\
Germany	&	0.39	&	0.47	&	0.13 &	0.01 & 2.05  \\
USA	&	0.28	&	0.49	&	0.19 &	0.04 & 2.5  \\
Israel	&	0.21	&	0.4	&	0.28 &	0.11 & 3.14  \\
Spain	&	0.19	&	0.53	&	0.26 &	0.02 & 2.69  \\
Norway	&	0.4	&	0.41	&	0.18 &	0.01 & 2.22  \\
Model	&	0.3	&	0.46	&	0.2 &	0.04 & 2.6  \\

  \bottomrule
\end{tabular}  \caption{\textbf{Household size distribution $P(m)$ per country}.\ For each country we show the fraction of households with $1$, $2-3$, $4-5$ or $6$ cohabitants, as obtained from the UN database \cite{si: UNHouseholds}. We also show the average household size in each country.}
  \label{table: household data by country}
\end{table}

\clearpage
\bibliographystyle{unsrt}


\end{document}